\documentclass[prd,amsmath,amssymb,aps,twocolumn]{revtex4-1}

\usepackage{subcaption}
\usepackage[dvipsnames]{xcolor}
\usepackage{graphicx}
\usepackage[colorlinks=true,citecolor=blue]{hyperref}
\usepackage{soul}
\setstcolor{Green}

\newcommand{\subE}{\textrm{\tiny{E}}}
\newcommand{\subS}{\textrm{\tiny{S}}}
\newcommand{\subHH}{\textrm{\tiny{HH}}}
\newcommand{\subDS}{\textrm{\tiny{DS}}}
  \newcommand{\be}{\begin{equation} }
 \newcommand{\ee}{\end{equation}}
    \newcommand{\bes}{\begin{equation*} }
 \newcommand{\ees}{\end{equation*}}
  \newcommand{\bea}{\begin{eqnarray} }
 \newcommand{\eea}{\end{eqnarray}}
    \newcommand{\beas}{\begin{eqnarray*} }
 \newcommand{\eeas}{\end{eqnarray*}}
   \newcommand{\ba}{\begin{align} }
 \newcommand{\ea}{\end{align} }
  \newcommand{\bas}{\begin{align*} }
   \newcommand{\eas}{\end{align*} }

\newcommand{\floor}[1]{\lfloor #1 \rfloor}

\usepackage{caption}

\begin{document}

\title{Vacuum polarization for varying quantum scalar field parameters in Schwarzschild-anti-de Sitter spacetime}
\author{Cormac Breen}
\email{cormac.breen@dit.ie}
\affiliation{School of Mathematical Sciences, Dublin Institute of Technology, Kevin Street, Dublin 8, Ireland}

\author{Peter Taylor}
\email{peter.taylor@dcu.ie}
\affiliation{
Center for Astrophysics and Relativity,School of Mathematical Sciences, Dublin City University, Glasnevin, Dublin 9, Irelandd}

\date{\today}
\begin{abstract}
 Equipped with  {new} powerful and efficient methods for computing quantum expectation values in static-spherically symmetric spacetimes in arbitrary dimensions, we  perform an in-depth investigation of how the { quantum} vacuum polarization varies with the parameters in the theory. In particular, we compute and compare the vacuum polarization for a quantum scalar field in the Schwarzschild anti-de Sitter black hole spacetime for a range of values of the field mass and field coupling constant  as well as the black hole mass and number of spacetime dimensions. In addition, a new approximation for the vacuum polarization in asymptotically anti-de Sitter black hole spacetimes is  presented.  \end{abstract}
\maketitle

\section{Introduction}
The expectation value of the quantum stress-energy tensor plays a crucial role in the semi-classical theory of gravity. It governs the quantum backreaction on the classical spacetime geometry via the semi-classical field equations
\begin{align}
	G_{ab}+\Lambda g_{ab}=8\pi \langle \hat{T}_{ab}\rangle.
\end{align}
Computing the expectation value of the quantum stress-energy tensor is beset with challenges. The main difficulty is that in quantizing the system, one promotes classical fields to operator-valued distributions, and since the classical stress-energy tensor is quadratic in the field, the quantum stress-energy tensor is quadratic in an operator-valued distribution, a mathematically ill-defined object. This implies that  $\langle \hat{T}_{ab}\rangle$ must be regularized. A conceptual framework for regularization, known as point-splitting regularization, was developed by DeWitt and Christensen \cite{DeWitt:1975, Christensen:1976}, { and put on a rigorous footing by Wald \cite{Wald:1977up}}. There still followed decades of effort towards a practical numerical implementation of point-splitting regularization in black hole spacetimes. This industry has been recently revived by new methods developed by Levi and Ori \cite{LeviOriPRL1} and applied to a scalar field on a Kerr black hole \cite{LeviOriPRL2}.

A related but technically less-challenging problem is that of computing the regularized vacuum polarization. It has become customary to apply techniques to this problem before applying to the stress-energy tensor, notwithstanding interest in the vacuum polarization in its own right, for example, in the phenomenon of spontaneous symmetry breaking (see e.g. \cite{Frolov:1998}). The vacuum polarization often shares important properties with the stress-energy tensor, e.g., they usually both diverge or are both regular on the horizon of the black hole for the field in a given quantum state.

In almost all calculations of the regularized stress-energy tensor or vacuum polarization in the literature, the emphasis is on describing the method for a given fixed set of parameters, mainly because these calculations are notoriously difficult and computationally expensive. However, the authors have devised a regularization method \cite{taylorbreen:2016,taylorbreen:2017}, which we will refer to as the extended coordinate method, which provides an extremely efficient way to compute the vacuum polarization for arbitrary field parameters in static spherically-symmetric spacetimes of arbitrary dimensions. In particular, this method allows us to explore the parameter space of the semi-classical theory in a way that would be so tedious as to be completely impractical using other schemes, for example, the Candelas Howard method \cite{CandelasHoward}.  

We are considering a scalar field satisfying the Klein-Gordon equation
\begin{align}
	\{\Box-m^{2}-\xi\,R\}\varphi (x)=0,
\end{align}
where $\Box$ is the d'Alembertian operator, $m$ is the field mass, $R$ is the Ricci curvature scalar of the background spacetime and $\xi$ is the coupling strength between the field and the background geometry. When considering vacuum polarization in a Ricci flat spacetime, the coupling constant is irrelevant (though not for the stress-energy tensor) since the $\xi\,R$ term vanishes in the wave equation. Hence, we choose to consider a quantum scalar field propagating on a black hole in anti-de Sitter spacetime,  namely, the Schwarzschild anti-de Sitter spacetime, which allows us to probe how the quantum effects vary with the coupling $\xi$ in a simple but non-trivial way, since the Ricci scalar on this background is constant but non-zero. As well as probing the dependence on the coupling constant, we compute the vacuum polarization for varying field mass, black hole mass and spacetime dimension. We note that, since the spacetime we are considering is not asymptotically flat, the black hole solution we consider here is not the only possibility. There are asymptotically anti-de Sitter solutions with other horizon topologies. We do not consider these cases here, but note that vacuum polarization for massless, conformally-coupled fields in these topological black hole spacetimes have been recently computed in Ref.~\cite{Morley:2018} by adapting the extended coordinate method of Ref.~\cite{taylorbreen:2016,taylorbreen:2017}.

This paper is organized as follows: In Section \ref{sec:review} we review the main features of the extended coordinate regularization method \cite{taylorbreen:2016,taylorbreen:2017}. In Section \ref{sec:results} we outline the main features of the Schwarzschild anti-de Sitter spacetime, discuss the numerical calculations required to construct the two-point function, present and discuss the main numerical results for the vacuum polarization and, finally in this section, we introduce and discuss a new approximation for the vacuum polarization. Lastly, in Section \ref{sec:con}, we draw some conclusions based upon the work presented in this paper.

\section{Review of Regularisation Method}
\label{sec:review}
Here we will briefly review the mode-sum regularization method of \cite{taylorbreen:2016,taylorbreen:2017}, we refer the reader to those papers for a comprehensive description.

We wish to compute the vacuum polarization for a scalar field in the Hartle-Hawking \cite{Hartle:1983} state in the Schwarzschild anti-de Sitter (which we will abbreviate to SadS henceforth) black hole spacetime. Since the spacetime is static, the vacuum polarization is conveniently defined in terms of a Euclideanized two-point function,
\begin{align}
	\label{eq:VPDef}
	\langle \hat{\phi}^{2}\rangle_{\subHH}=\lim_{x'\to x} \left\{G_{\subE}(x,x')-G_{\subS}(x,x')\right\}
\end{align}
where $G_{\subE}(x,x')$ a Green function for the wave equation on the Euclidean black hole spacetime obtained by a Wick rotation $t\to-i\tau$, and $G_{\subS}(x,x')$ is a symmetric two-point function which is a parametrix for the wave operator and which is constructed only from the geometry of the spacetime through its metric and its derivatives \cite{Wald:1995yp}. In particular, we take $G_{\subS}(x,x')$ to be a Hadamard parametrix of the form,
\begin{align}
\label{eq:Had}
G_{\subS}(x,x')=&\frac{\Gamma(\frac{d}{2}-1)}{2 (2\pi)^{d/2}}\Bigg\{\frac{U(x,x')}{\sigma(x,x')^{\frac{d}{2}-1}}\nonumber\\
&+V(x,x')\log(2 \sigma(x,x')/\ell^2)\Bigg\}.
\end{align}
The biscalar $\sigma(x,x')$ is the world function with respect to the Euclideanized metric. The parameter $\ell$ is an arbitrary length scale required to make the argument of the log dimensionless.  The biscalars $U(x,x')$ and $V(x,x')$ are smooth and symmetric in their arguments. For even $d$, $V(x,x')$ is a homogeneous solution of the wave equation in both its arguments; for odd $d$, $V(x,x')\equiv 0$. High order covariant Taylor expansions for these biscalar can be found in \cite{DecaniniFolacci:2008} . By construction, the difference in (\ref{eq:VPDef}) is finite in the coincidence limit $x'\to x$, albeit difficult to compute in practice.

To see where the difficulty lies, we note that $G_{\subE}(x,x')$ is not known in closed form, but can only be expressed as a mode-sum, which we derive later in (\ref{eq:Gmodesum})-(\ref{eq:radialeqn}). The divergences as $x'\to x$ in this mode-sum manifest in the fact that the sums do not converge in this limit, while in $G_{\subS}(x,x')$, they are explicitly geometrical. The mode-sum approach developed in \cite{taylorbreen:2016,taylorbreen:2017} involves a Fourier and multipole decomposition of the Hadamard parametrix so that the difference in (\ref{eq:VPDef}) can be taken mode-by-mode. Before describing this, we first derive the Euclidean Green function for a scalar field in the SadS black hole spacetime.

\subsection{The Euclidean Green Function}
The SadS black hole spacetime is a static, spherically-symmetric solution to the vacuum Einstein equations
\begin{align}
	G_{ab}-\Lambda g_{ab}=0,
\end{align}
with a negative cosmological constant $\Lambda<0$. In Schwarzschild-like coordinates, the Euclideanized version of this solution has a line-element of the form 
\begin{align}
\label{eq:metric}
ds^{2}=f(r)d\tau^{2}+dr^{2}/f(r)+r^{2}d\Omega^{2}_{d-2},
\end{align}
where $d\Omega^{2}_{d-2}$ is the metric on $\mathbb{S}^{d-2}$ and
\begin{align}
\label{eq:fSadS}
f(r)=1-\frac{\varpi_{d}}{r^{d-3}}+\frac{r^2}{L^2},
\end{align}
where $\varpi_{d}$ is the mass parameter related to the conserved mass $M$ by \cite{Ashtekar:1999jx}
\begin{align}
\label{eq:mass}
    M=\frac{(d-2)\Omega_{d-2}\varpi_{d}}{16\pi},\quad \Omega_{d-2}=\frac{2\,\pi^{(d-1)/2}}{\Gamma(\frac{d-1}{2})},
\end{align}
and
\begin{align}
    L=\sqrt{\frac{-(d-1)(d-2)}{2\Lambda}}
\end{align}
is the adS curvature lengthscale. 
The Ricci Scalar for SadS { in $d$-dimensional spacetime} is given by:
\begin{align}
\label{eq:Ricci}
R=-\frac{d(d-1)}{L^2}.
\end{align}
These coordinates are singular when $f(r)=0$, which corresponds to an horizon. When $d$ is even, $f(r)$ has a single real root $r=r_{\textrm{h}}$ corresponding to a black hole horizon. When $d$ is odd, $f(r)$ has two real roots $r=\pm r_{\textrm{h}}$, of which the positive root corresponds to a black hole horizon. For example,
\begin{align}
r_{\textrm{h}}&= \left(\frac{\Delta_4}{9}\right)^{1/3}-\frac{L^2}{(3 \Delta_4)^{1/3}} & &\mbox{for } d=4\\
r_{\textrm{h}}&= \sqrt{\frac{\Delta_5}{2}} & &\mbox{for } d=5\
\end{align}
with $\Delta_4=\tfrac{9}{2} L^2 \left(\varpi_{4}+ \sqrt{\tfrac{12}{81}L^2+ \varpi_{4}^2}\right)$ and $\Delta_5=L \left(\sqrt{L^2+4 \varpi_{5}}-L\right)$.

It can be shown that the Euclidean metric would possess a conical singularity on the horizon unless we enforce the periodicity $\tau=\tau+2\pi/\kappa$ where $\kappa$ is the surface gravity. Imposing this periodicity discretizes the frequency spectrum of the field modes which now satisfy an elliptic wave equation
\begin{align}
(\Box_{\subE}-m^2-\xi\,R)\phi=0,
\end{align}
where $\Box_{\subE}$ is the d'Alembertian operator with respect to the Euclidean metric. The corresponding Euclidean Green function has the following mode-sum representation
\begin{align}
	\label{eq:Gmodesum}
G_{\subS}(x,x')=\frac{\kappa}{2\pi}\sum_{n=-\infty}^{\infty}e^{i n \kappa \Delta\tau}\sum_{l=0}^{\infty}\frac{(2l+2\mu)}{\Omega_{d-2}}C_{l}^{\mu}(\cos\gamma)g_{nl}(r,r')
\end{align}
where $\mu=(d-3)/2$, $C_{l}^{\mu}(x)$ is the Gegenbauer polynomial and $\gamma$ is the geodesic distance between two points on the $(d-2)$-sphere. The radial Green function satisfies
\begin{align}
\label{eq:radialeqn}
\Bigg[\frac{d}{dr}\Big(r^{d-2}f(r)\frac{d}{dr}\Big)-r^{d-4}l(l+d-3)-r^{d-2}\frac{n^{2}\kappa^{2}}{f(r)}\nonumber\\
-r^{d-2}(m^{2}+\xi\,R)\Bigg]g_{nl}(r,r')=-\delta(r-r').
\end{align}
The solution can be expressed as a normalized product of homogeneous solutions
\begin{align}
\label{eq:radialgreenfn}
g_{nl}(r,r')=N_{nl}\,p_{nl}(r_{<})q_{nl}(r_{>}),
\end{align}
where $p_{nl}(r)$ and $q_{nl}(r)$ are homogeneous solutions which are regular on the horizon and the outer boundary (usually spatial infinity), respectively. We have adopted the notation $r_{<}\equiv \min\{r,r'\}$, $r_{>}\equiv \max\{r,r'\}$. The normalization constant is given by
\begin{align}
\label{eq:norm}
N_{nl}=-r^{d-2}f(r)\,W\{p_{nl}(r),q_{nl}(r)\},
\end{align}
where $W\{p,q\}$ denotes the Wronskian of the two solutions.

\subsection{Mode-Sum Representation of the Hadamard Parametrix}
In order to compute the limit in (\ref{eq:VPDef}), we need to express (\ref{eq:Had}) in the same set of basis modes as the Green function (\ref{eq:Gmodesum}). Following Ref. \cite{taylorbreen:2016,taylorbreen:2017}, we simplify by taking the partial coincidence limit $r'\to r$, and then rather than expanding the Hadamard parametrix in coordinate separation $\Delta x=x-x'$, we separate in so-called extended coordinates
\begin{align}
w^{2}=\frac{2}{\kappa^{2}}(1-\cos \kappa\Delta\tau),\qquad s^{2}=f(r)\,w^{2}+2 r^{2}(1-\cos\gamma).
\end{align}
The extended coordinates $w$ and $s$ are formally treated as $\textrm{O}(\epsilon)\sim\textrm{O}(\Delta x)$ quantities. Then, for example, the direct part of the Hadamard parametrix possesses an expansion of the form
\begin{align}
\frac{U}{\sigma^{\frac{d}{2}-1}}=\sum_{i=0}^{ \floor{\frac{d+1}{2}}}\sum_{j=0}^{ i}\mathcal{D}^{(+)}_{ij}(r)\epsilon^{2i-2\mu-1}\frac{w^{2i+2j}}{s^{2\mu+2j+1}}\nonumber\\
+\sum_{i=0}^{ \floor{\frac{d+1}{2}}}\sum_{j=1}^{ i}\mathcal{D}^{(-)}_{ij}(r)\epsilon^{2i-2\mu-1}\frac{w^{2i-2j}}{s^{2\mu-2j+1}}+\textrm{O}(\epsilon^{4}),
\end{align}
where the $\mathcal{D}^{(\pm)}_{ij}(r)$ coefficients are too lengthy to print, particularly for higher numbers of dimensions, but a Mathematica Notebook with the explicit coefficients can be found online \cite{peterswebsite}. A similar expansion results for the tail term, up to the order being considered here, we have
\begin{align}
	V\log(2\sigma/\ell^{2})&=\log(\epsilon^{2}\,s^{2}/\ell^{2})\sum_{i=0}^{1}\sum_{j=0}^{i}\mathcal{T}^{\textrm{(l)}}_{ij}(r)\epsilon^{2i}s^{2i-2j}w^{2j}\nonumber\\
	&+\sum_{j=0}^{1}\mathcal{T}^{\textrm{(p)}}_{1j}(r)\epsilon^{2}s^{2-2j}w^{2j}+\mathcal{T}^{\textrm{(r)}}_{10}(r)\epsilon^{2}s^{-2}w^{4}\nonumber\\
	&+\mathcal{O}(\epsilon^{4}\log \epsilon)
\end{align}
where as before the tail coefficients $\mathcal{T}_{ij}$ can be found online \cite{peterswebsite}; they vanish identically for odd $d$ since $V\equiv 0$. Appropriate mode-sum representations of these expressions are obtained by expanding the $w$ and $s$-dependent parts in the mode functions used to expand the Green function, for example,
\begin{align}
\frac{w^{2i\pm 2j}}{s^{2\mu\pm 2j+1}}=\sum_{n=-\infty}^{\infty}e^{in\kappa\Delta\tau}\sum_{l=0}^{\infty}(2 l+2\mu)C_{l}^{\mu}(\cos\gamma)\nonumber\\
\times\,\,\stackrel{[d]}{\Psi}\!\!{}_{nl}(i,\pm j|r),
\end{align}
where the $\stackrel{[d]}{\Psi}\!\!{}_{nl}(i,\pm j|r)$ are known as regularization parameters, which are determined by inverting the expression above. Remarkably, in Refs.~\cite{taylorbreen:2016,taylorbreen:2017}, these regularization parameters have been derived in closed form for arbitrary numbers of dimensions and for arbitrary metric function $f(r)$. Analogous statements hold for the regularization parameters for the term involving the logarithm. The main result of Refs.~\cite{taylorbreen:2016,taylorbreen:2017} can now be stated as follows: For odd $d$, the Hadamard parametrix for a scalar field in a static, spherically symmetric spacetime has the mode-sum representation,
\begin{widetext}
\begin{align}
G_{\subS}(x,x')=\frac{\Gamma(\frac{d}{2}-1)}{2(2\pi)^{d/2}}\sum_{l=0}^{\infty}(2l+2\mu)C_{l}^{\mu}(\cos\gamma)\sum_{n=-\infty}^{\infty}e^{in\kappa\Delta\tau}\Big\{\sum_{i=0}^{\frac{d-1}{2}}\sum_{j=0}^{i}\mathcal{D}^{(+)}_{ij}(r)\stackrel{[d]}{\Psi}\!\!{}^{(+)}_{nl}(i,j|r)+\sum_{i=1}^{\frac{d-1}{2}}\sum_{j=1}^{i}\mathcal{D}^{(-)}_{ij}(r)\stackrel{[d]}{\Psi}\!\!{}^{(-)}_{nl}(i,j|r)\Big\}
\end{align}
where the regularization parameters are
	\begin{align}
	\label{eq:RegParamP}
	\stackrel{[d]}{\Psi}\!\!{}^{(+)}_{nl}(i,j|r)&=\frac{2^{2i-j-1}(-1)^{n}i!\,\Gamma(i+\frac{1}{2})\Gamma(\mu)}{\pi\kappa^{2i + 2j}r^{2\mu+ 2j+1}\Gamma(j+\mu+\tfrac{1}{2})}\Big(\frac{1}{\eta}\frac{d}{d \eta}\Big)^{j}\Big\{\frac{P_{l+\mu-\frac{1}{2}}(\eta)Q_{l+\mu-\frac{1}{2}}(\eta)}{(i-n)!(i+n)!}\nonumber\\
	&+\sum_{k=\max\{1,n-i\}}^{i+n}\frac{P^{-k}_{l+\mu-\frac{1}{2}}(\eta)Q^{k}_{l+\mu-\frac{1}{2}}(\eta)}{(i+k-n)!(i-k+n)!}+\sum_{k=\max\{1,-n-i\}}^{i-n}\frac{P^{-k}_{l+\mu-\frac{1}{2}}(\eta)Q^{k}_{l+\mu-\frac{1}{2}}(\eta)}{(i+k+n)!(i-k-n)!}\Big\}\nonumber\\
\stackrel{[d]}{\Psi}\!\!{}^{(-)}_{nl}(i,j|r)&=\frac{2^{2i-2j-1}(-1)^{n+j}(i-j)!\Gamma(i-j+\tfrac{1}{2})\Gamma(\mu)}{\pi\kappa^{2i - 2j}r^{2\mu- 2j+1}\Gamma(\mu+\tfrac{1}{2}-j)}\sum_{k=0}^{j}(-1)^k\binom{j}{k}\frac{l+\mu+j-2k}{(l+\mu-k)_{j+1}}\nonumber\\
	&\times\Big\{\frac{P_{l+\mu-\frac{1}{2}+j-2k}(\eta)Q_{l+\mu-\frac{1}{2}+j-2k}(\eta)}{(i-j-n)!(i-j+n)!}+\sum_{p=\max\{1,n-i+j\}}^{i-j+n}\frac{P^{-p}_{l+\mu-\frac{1}{2}+j-2k}(\eta)Q^{p}_{l+\mu-\frac{1}{2}+j-2k}(\eta)}{(i-j+p-n)!(i-j-p+n)!}\nonumber\\
&+\sum_{p=\max\{1,-n-i+j\}}^{i-j-n}\frac{P^{-p}_{l+\mu-\frac{1}{2}+j-2k}(\eta)Q^{p}_{l+\mu-\frac{1}{2}+j-2k}(\eta)}{(i-j+p+n)!(i-j-p-n)!}\Big\},\nonumber\\
	\end{align}
with 
\begin{align}
	\eta\equiv\sqrt{1-\frac{f(r)}{\kappa^{2}r^{2}}},
\end{align} 
$(z)_{\nu}$ is the Pochhamer symbol and $P^{\mu}_{\nu}(z)$, $Q^{\mu}_{\nu}(z)$ are the associated Legendre functions of the first and second kind, respectively, with the branch cut chosen to be on  $(-\infty,1]$. For even $d$, the mode-sum representation of the Hadamard parametrix is even more complicated as a result of the tail terms, the result is
\begin{align}
G_{\subS}&(x,x')=\frac{\Gamma(\frac{d}{2}-1)}{2(2\pi)^{d/2}}\sum_{l=0}^{\infty}(2l+2\mu)C_{l}^{\mu}(\cos\gamma)\sum_{n=-\infty}^{\infty}e^{in\kappa\Delta\tau}\Bigg\{\sum_{i=0}^{\frac{d}{2}}\sum_{j=0}^{i}\mathcal{D}^{(+)}_{ij}(r)\stackrel{[d]}{\Psi}\!\!{}^{(+)}_{nl}(i,j|r)\nonumber\\
&+\sum_{i=1}^{\frac{d}{2}}\sum_{j=1}^{\min[i,\frac{d}{2}-2]}\mathcal{D}^{(-)}_{i,j}(r)\stackrel{[d]}{\Psi}\!\!{}^{(-)}_{nl}(i,j|r)+\mathcal{T}^{(r)}_{10} \stackrel{[d]}{\Psi}\!\!{}^{(-)}_{nl}(\tfrac{d}{2},\tfrac{d}{2}-2|r)+\sum_{i=0}^{2}\sum_{j=0}^{i}\mathcal{T}^{(l)}_{ij} \stackrel{[d]}{\chi}\!\!{}_{nl}(i,j|r)\Bigg\}+\frac{\Gamma(\frac{d}{2}-1)}{2(2\pi)^{d/2}}\mathcal{D}^{(-)}_{\tfrac{d}{2}-1,\tfrac{d}{2}-1}(r)
\end{align}
where the $\stackrel{[d]}{\Psi}\!\!{}^{(\pm)}_{nl}(i,j|r)$ are given by (\ref{eq:RegParamP}) and the $\stackrel{[d]}{\chi}\!\!{}_{nl}(i,j|r)$ are
	\begin{align}
\stackrel{[d]}{\chi}\!\!{}_{nl}(i,j|r)=(-1)^{n+\mu-\tfrac{1}{2}}\frac{2^{2j-1}(i-j)!j!\Gamma(j+\tfrac{1}{2})\Gamma(\mu)}{\pi\kappa^{2j}r^{2j-2i}}\sum_{k=0}^{\mu+\tfrac{1}{2}+i-j}(-1)^k\binom{\mu+\tfrac{1}{2}+i-j}{k}\frac{l+2\mu+i-j-2k+\tfrac{1}{2}}{(l+\mu-k)_{\mu+\tfrac{3}{2}+i-j}}&\nonumber\\
\times\Big\{\frac{P_{l+2\mu+i-j-2k}(\eta)Q_{l+2\mu+i-j-2k}(\eta)}{(j-n)!(j+n)!}+\sum_{p=\max\{1,n-j\}}^{j+n}\frac{P^{-p}_{l+2\mu+i-j-2k}(\eta)Q^{p}_{l+2\mu+i-j-2k}(\eta)}{(j+p-n)!(j-p+n)!}&\nonumber\\
+\sum_{p=\max\{1,-n-j\}}^{j-n}\frac{P^{-p}_{l+2\mu+i-j-2k}(\eta)Q^{p}_{l+2\mu+i-j-2k}(\eta)}{(j+p+n)!(j-p-n)!}\Bigg\}&\nonumber\\
\mbox{for } l>i-j&\nonumber\\
	\end{align}
and
	\begin{align}
	\label{eq:tailnum}
	\stackrel{[d]}{\chi}\!\!{}_{nl}(i,j|r)=\frac{\kappa}{(2\pi)^2}\sqrt{\pi}2^{\mu +\frac{1}{2}} \Gamma(\mu)(2r^2)^{i-j}(-1)^l \left(\frac{2}{\kappa^2}\right)^j\Bigg[\frac{d}{d\lambda} (\lambda+1-l)_l \left(\frac{2r^2}{\ell^2}\right)^{\lambda-i+j} \int_{0}^{2\pi/\kappa}(1-\cos \kappa t)^je^{-i n \kappa t}\nonumber\\
	\times ~ (z^2-1)^{\frac{1}{2}(\mu+\lambda+\frac{1}{2})}\mathcal{Q}_{l+\mu-\frac{1}{2}}^{-\mu-\lambda-\frac{1}{2}} d t\Bigg]_{\lambda=i-j} ~~~ \mbox{for } l\leq i-j.
	\end{align}
\end{widetext}
In the last expression, $\mathcal{Q}^{\nu}_{\mu}(z)$ is Olver's definition of the Legendre function of the second kind \cite{Olver}, valid for all values of $\mu$, $\nu$.

\section{Numerical Results}
\label{sec:results}
\subsection{Radial Modes for the Euclidean Green Function}
We begin this section by describing the numerical computation of the radial modes $p_{nl}(r)$ and $q_{nl}(r)$, the homogeneous solutions to (\ref{eq:radialeqn}) needed for the Euclidean Green function, which are regular (for all values of the field parameters) on the horizon and at infinity, respectively. For $f(r)$ given by (\ref{eq:fSadS}), solutions cannot be given in terms of known functions and must be solved numerically.

For the SadS spacetime, Eq.~(\ref{eq:radialeqn}) has regular singular points at both $r=r_{\textrm{h}}$ and at infinity. This is in contrast to the asymptotically flat case where infinity is an irregular singular point. That both $r=r_{\textrm{h}}$ and infinity are regular singular points implies that Eq.~ (\ref{eq:radialeqn}) admits the following Frobenius series solutions
\begin{align}
P_{nl}{ (r)}&=\sum_{i=0}^{\infty}a_i (r-r_{\textrm{h}})^{i+\alpha} && \text{about  } r=r_{\textrm{h}}\label{eq:pseries}\\
Q_{nl}{ (r)}&=\sum_{i=0}^{\infty}b_i \left(\frac{1}{r}\right)^{i+\rho} && \text{about  }r=\infty\label{eq:qseries}
\end{align}
with $\alpha$, $\rho$ the indicial exponents to be determined. Inserting these series into (\ref{eq:radialeqn}) yields, for the indicial exponents,
\begin{align}
\alpha_{\pm} &=\pm \frac{|n|}{2}\\
\label{eq:indicial}
\rho_{\pm} &= \frac{d-1}{2}\pm \sqrt{\mu_{\xi}+\frac{1}{4}},
\end{align}
where we have found it convenient to define the dimensionless effective field mass
\begin{align}
	\label{eq:muxi}
	\mu_{\xi}\equiv L^{2}\left(m^{2}+\left(\xi-\xi_{c}\right)R\right),\quad \xi_{c}=\frac{1}{4}\left(\frac{d-2}{d-1}\right),
\end{align}
and we have used the explicit expression for the Ricci scalar (\ref{eq:Ricci}) to arrive at (\ref{eq:indicial}). With this definition, the redefined field mass vanishes for massless, conformally coupled fields and the solutions to the indicial equation in this case are simply $\rho=d/2$ or $\rho=d/2-1$. Requiring that $\rho$ is real gives $\mu_{\xi}\ge -1/4$, which we can re-express as an upper bound on $\xi$,
\begin{align}
	\label{eq:ximax}
\xi \leq \xi_u = \frac{m^2 L^{2}}{d(d-1)}+ \frac{d-1}{4 d}. 
\end{align}
This is completely analogous to the Breitenlohner-Freedman bound for a massive Klein-Gordon field in adS \cite{Breitenlohner:1982bm}.

Now, for the quantum field to be in the Hartle-Hawking state requires that $p_{nl}(r)$ be the solution regular at the horizon, so we choose the following leading order behaviour,
\begin{align}
p_{nl}(r) &\sim (r-r_{\textrm{h}})^{\frac{|n|}{2}} &&r \to r_{\textrm{h}}.
 \label{eq:pinf}
\end{align}
Unlike the asymptotically flat case, the spacetime is not globally hyperbolic and this implies that we must also specify boundary conditions at the timelike boundary $r=\infty$ in order for the Klein Gordon equation to be well-posed \cite{Vasy}. It has been shown \cite{Holzegel:2011qj, Vasy} that the wave equation under consideration has a well-posed initial value formulation for fields satisfying the Breitenlohner-Freedman bound (\ref{eq:ximax}) with Dirichlet boundary conditions. This corresponds to taking $q_{nl}(r)$ to be the solution satisfying the fast fall-off condition (corresponding to the indicial exponent $\rho_{+}$) at infinity
\begin{align}
	\label{eq:qasymp}
q_{nl}(r) &\sim r^{-\frac{d-1}{2} -  \sqrt{ \mu_{\xi}+\frac{1}{4}}} & &r \to \infty.
\end{align}
The radial mode solutions are invariant under $n\to-n$ so we need only consider positive frequency modes. To compute $p_{nl}(r)$, we integrate the homogeneous version of (\ref{eq:radialeqn}) from an initial point near the horizon outwards using the Mathematica routine {\tt NDSolve}. A high-order Frobenius series of the form (\ref{eq:pseries}) is obtained and used, along with its first derivative, as initial data for the integration. In the numerical implementation, the internal working precision of each calculation is set to 60 digits while the accuracy and precision goals (i.e. the effective number of digits of precision and accuracy sought in the final result) were both set to 35 digits.

Similarly we obtain $q_{nl}(r)$ by integrating the homogeneous version of (\ref{eq:radialeqn}) inwards from some large $r$-value with initial data generated from the high order Frobenius series (\ref{eq:qseries}), again using {\tt NDSolve} with the same accuracy and precision as before.
 As a check of the accuracy of the calculated modes, we use the modes to compute the Wronskian and hence the normalization in Eq.~(\ref{eq:norm}) at all $r$-values across the domain of integration and for all values of the parameters $l$ and $n$. We find that $N_{nl}$ remains constant in $r$ to within $10^{-30} |N_{nl}|$ for all $l$ and $n$, confirming the accuracy of the numerical calculation of the modes. 

\subsection{Mode-Sum Vacuum Polarization}
Equipped with an accurate numerical evaluation of the radial Green function and a mode-sum representation of the Hadamard parametrix from \cite{taylorbreen:2016,taylorbreen:2017}, we are now in a position to calculate the vacuum polarization $\langle \hat{\phi}^2 \rangle_{\subHH}$ for a scalar field in the Hartle-Hawking state in an SadS spacetime. Using the $n\to-n$ symmetry and the well-known values of the Gegenbauer polynomials when the argument is unity, we can express the vacuum polarization as
\begin{align}
\langle \hat{\phi}^{2} \rangle_{\subHH}&=\frac{\kappa}{2\pi}\sum_{l=0}^{\infty}\frac{(2l+2\mu)}{\Omega_{d-2}}\binom{2\mu+l-1}{l}\Big\{g_{0l}(r)-g_{0l}^{\subS}(r)\nonumber\\
&	+2\sum_{n=1}^{\infty}(g_{nl}(r)-g_{nl}^{\subS}(r))\Big\}\nonumber\\
	g_{nl}^{\subS}(r)=&\frac{\Gamma(\frac{d}{2}-1)}{2(2\pi)^{d/2}}\Big\{\sum_{i=0}^{\frac{d-1}{2}}\sum_{j=0}^{i}\mathcal{D}^{(+)}_{ij}(r)\stackrel{[d]}{\Psi}\!\!{}^{(+)}_{nl}(i,j|r)\nonumber\\
	&+\sum_{i=1}^{\frac{d-1}{2}}\sum_{j=1}^{i}\mathcal{D}^{(-)}_{ij}(r)\stackrel{[d]}{\Psi}\!\!{}^{(-)}_{nl}(i,j|r)\Big\},
\end{align}
in the odd dimensional case and
\begin{align}
\langle \hat{\phi}^{2} \rangle_{\subHH}&=\frac{\kappa}{2\pi}\sum_{l=0}^{\infty}\frac{(2l+2\mu)}{\Omega_{d-2}}\binom{2\mu+l-1}{l}\Big\{g_{0l}(r)-g_{0l}^{\subS}(r)\nonumber\\
&+2\sum_{n=1}^{\infty}(g_{nl}(r)-g_{nl}^{\subS}(r))\Big\}+ \frac{\Gamma(\frac{d}{2}-1)}{2(2\pi)^{d/2}}\mathcal{D}^{(-)}_{\tfrac{d}{2}-1,\tfrac{d}{2}-1}(r)\nonumber\\
g_{nl}^{\subS}(r)=&\frac{\Gamma(\frac{d}{2}-1)}{2(2\pi)^{d/2}}\Big\{\sum_{i=0}^{\frac{d}{2}}\sum_{j=0}^{i}\mathcal{D}^{(+)}_{ij}(r)\stackrel{[d]}{\Psi}\!\!{}^{(+)}_{nl}(i,j|r)\nonumber\\
&+\sum_{i=1}^{\frac{d}{2}}\sum_{j=1}^{\min[i,\frac{d}{2}-2]}\mathcal{D}^{(-)}_{ij}(r)\stackrel{[d]}{\Psi}\!\!{}^{(-)}_{nl}(i,j|r)\Big\}\nonumber\\
&+\mathcal{T}^{(r)}_{10} \stackrel{[d]}{\Psi}\!\!{}^{(-)}_{nl}(\tfrac{d}{2},\tfrac{d}{2}-2|r)+\sum_{i=0}^{2}\sum_{j=0}^{i}\mathcal{T}^{(l)}_{ij} \stackrel{[d]}{\chi}\!\!{}_{nl}(i,j|r)\Big\},
\end{align}
in the even dimensional case.

Each of these mode-sum expressions for $\langle \hat{\phi}^{2} \rangle_{\subHH}$ are rapidly convergent and can be truncated at a modest finite frequency and multipole cut-off yielding very accurate results for the vacuum polarization. In the plots shown and discussed below, we truncated the frequency sum at $n_{\textrm{max}}=10$ and the multipole sum at $l_{\textrm{max}}=70$. This enables  $\langle \hat{\phi}^{2} \rangle_{\subHH}$ to be calculated on a standard laptop in a reasonably short time, less than 2 hours for $d=4$ for example. Our answers are accurate to at least 7 decimal places everywhere on the exterior of the black hole.  Since it is only the relative sizes of the adS lengthscale $L$ compared with the black hole lengthscale $M$ that is relevant, we fix the adS lengthscale to be unity in all the results that follow. We further fix the arbitrary lengthscale that appears in the Hadamard parametrix to be unity so that $L=\ell=1$.

\subsection{Results and Discussion}

\subsubsection{Varying $m$}
\label{sec:m}
\begin{figure*} 
	\begin{subfigure}{.5\textwidth}
		\includegraphics[width=.8\linewidth]{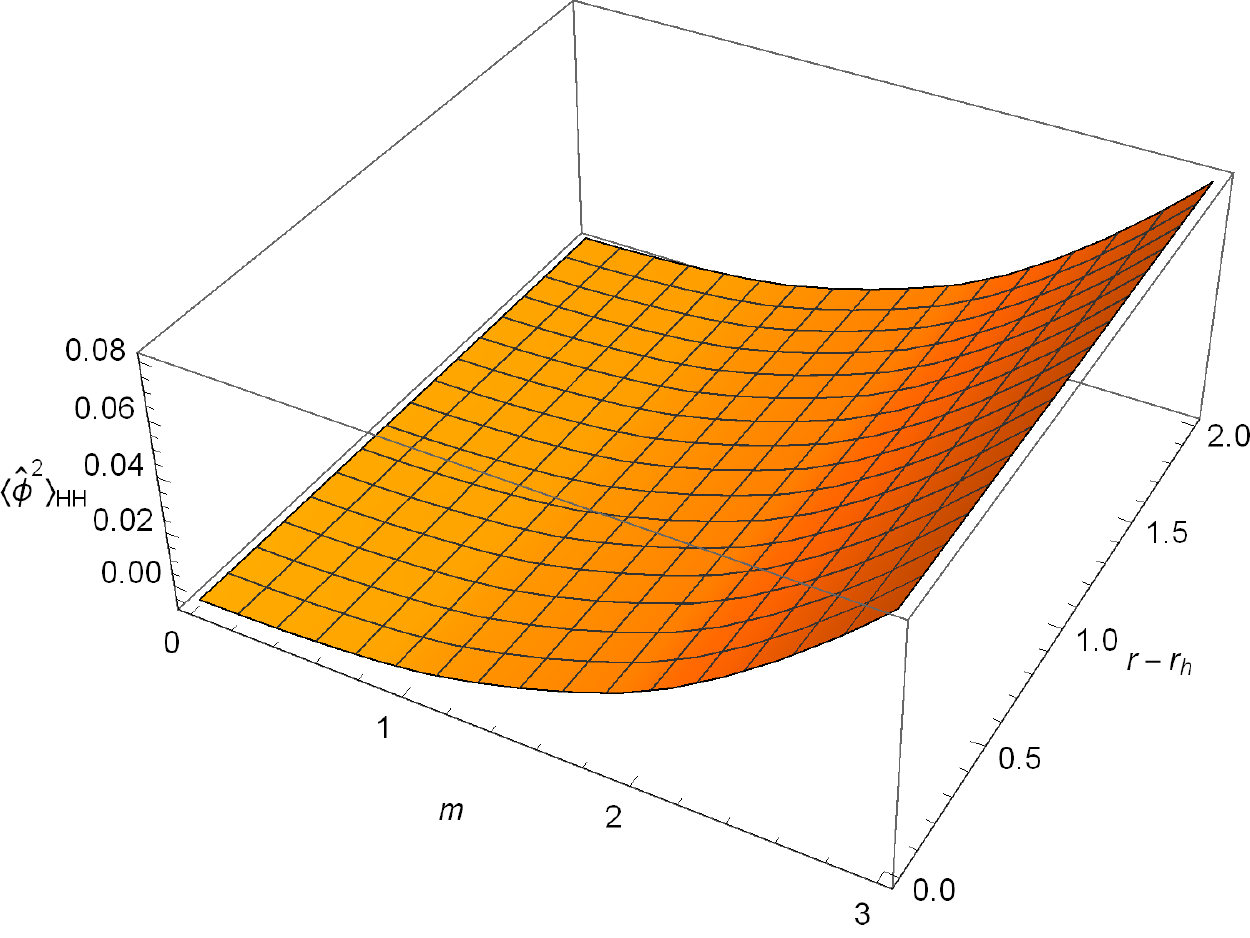}
		\caption{$d=4,\xi=0$}
		\label{fig:d4ofmxi0}
	\end{subfigure}%
	\begin{subfigure}{.5\textwidth}
	\includegraphics[width=.8\linewidth]{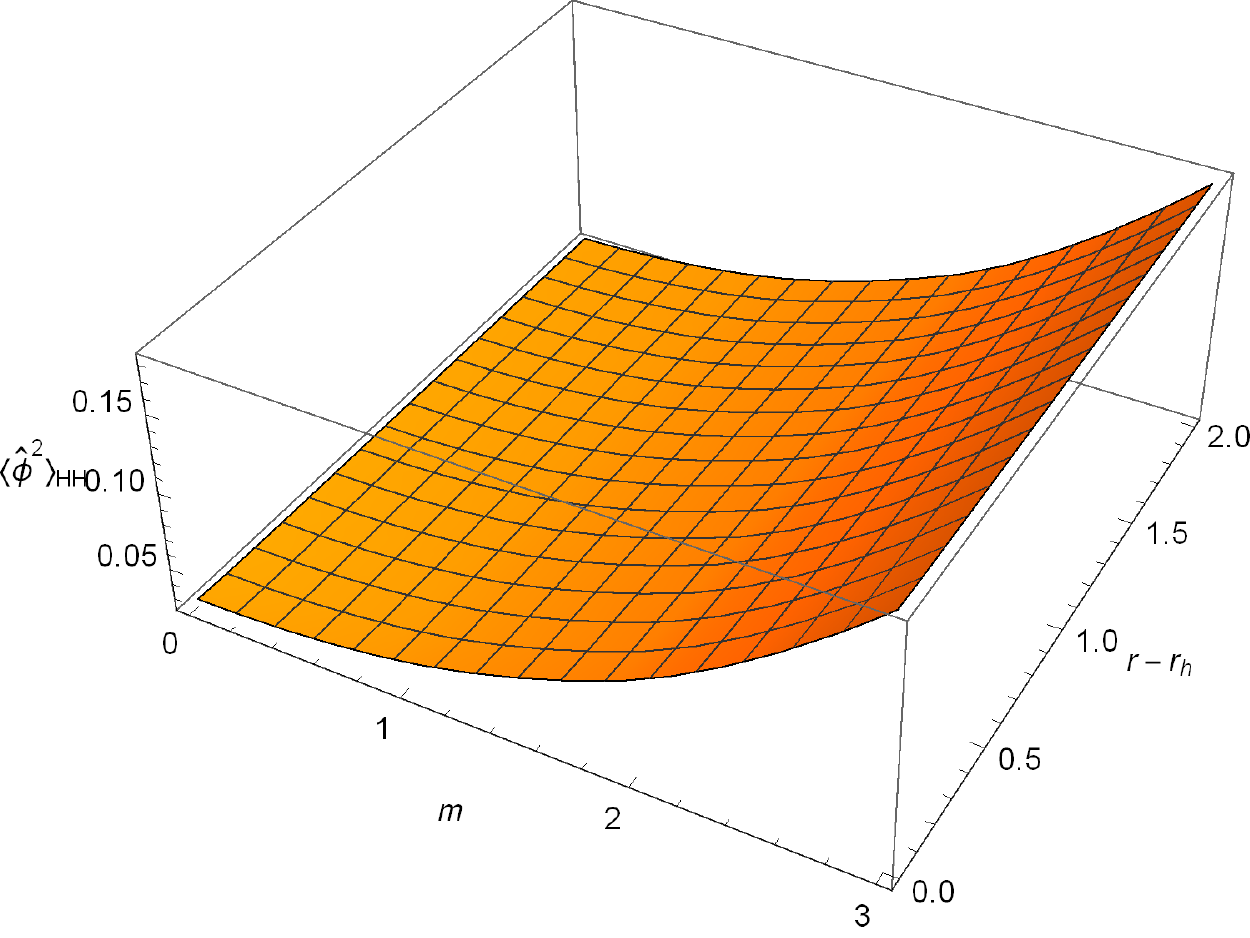}
	\caption{$d=5,\xi=0$}
	\label{fig:d5ofmxi0}
\end{subfigure}
	\begin{subfigure}{.5\textwidth}
	\includegraphics[width=.8\linewidth]{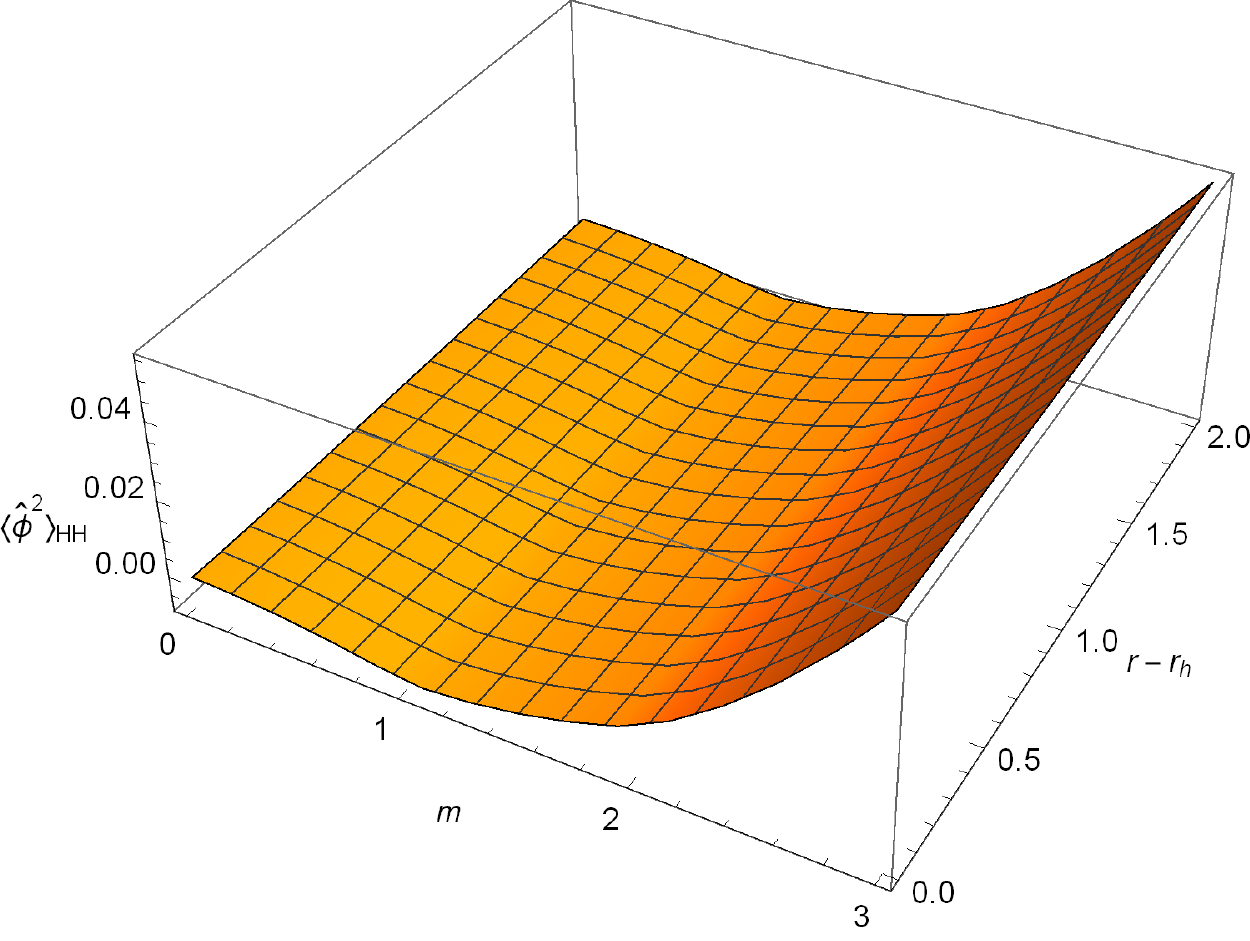}
	\caption{$d=4,\xi=1/6$}
	\label{fig:d4ofmxi16}
\end{subfigure}%
\begin{subfigure}{.5\textwidth}
	\includegraphics[width=.8\linewidth]{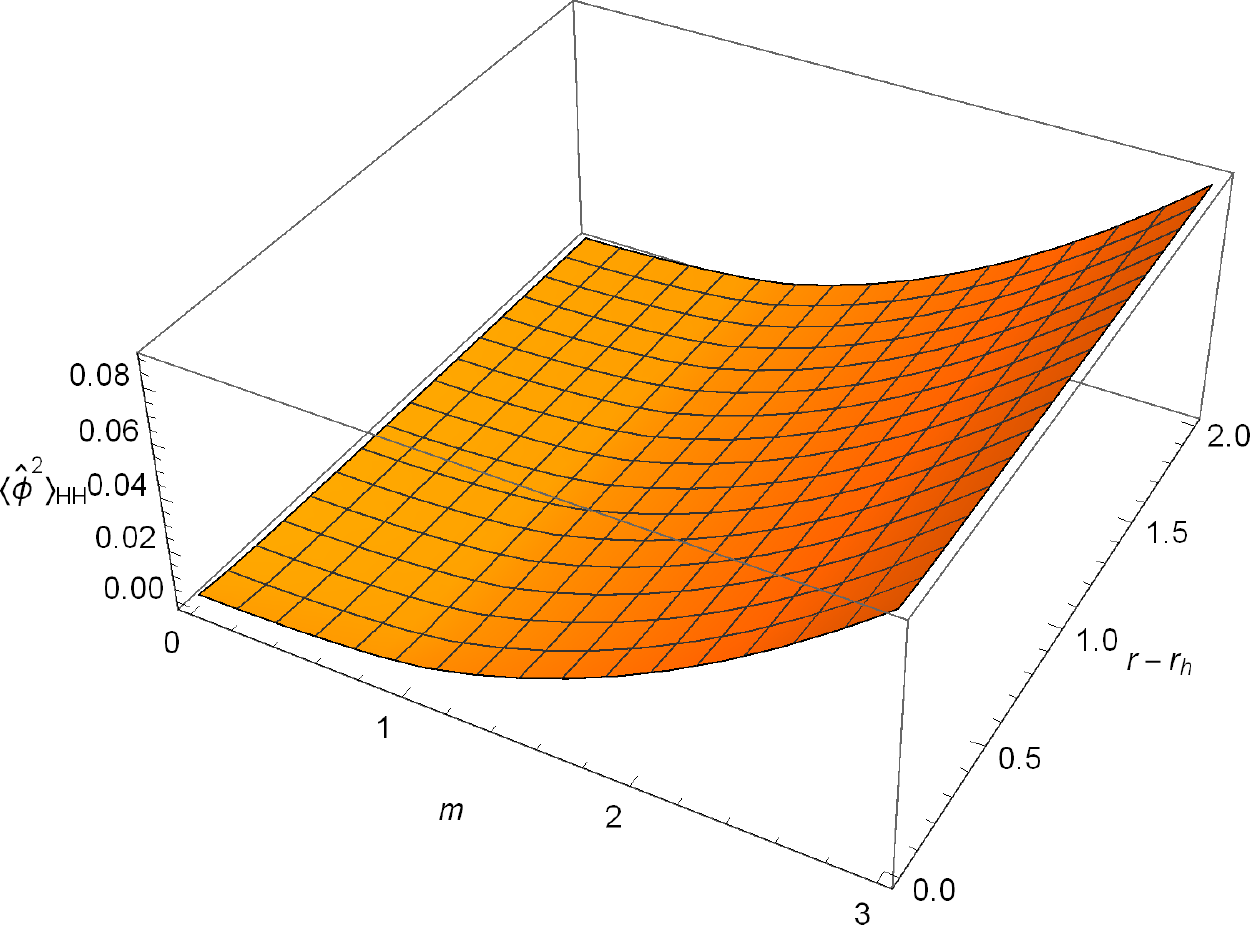}
	\caption{$d=5,\xi=1/6$}
	\label{fig:d5ofmxi16}
\end{subfigure}%
	\caption{\label{fig:m1} Plots of the vacuum polarization as a function of the field mass $m$ and distance from the black hole for a scalar field in SadS spacetime in $d=4$ and $d=5$ dimensions, and with coupling constant $\xi=0$ and $\xi=1/6$. The black hole mass parameter $\varpi_{d}$ has been set to twice the adS lengthscale so that $\varpi_{d}=2\,L=2\,\ell=2$.}	
\end{figure*}
\begin{figure*}
	\begin{subfigure}{.5\textwidth}
		\includegraphics[width=.8\linewidth]{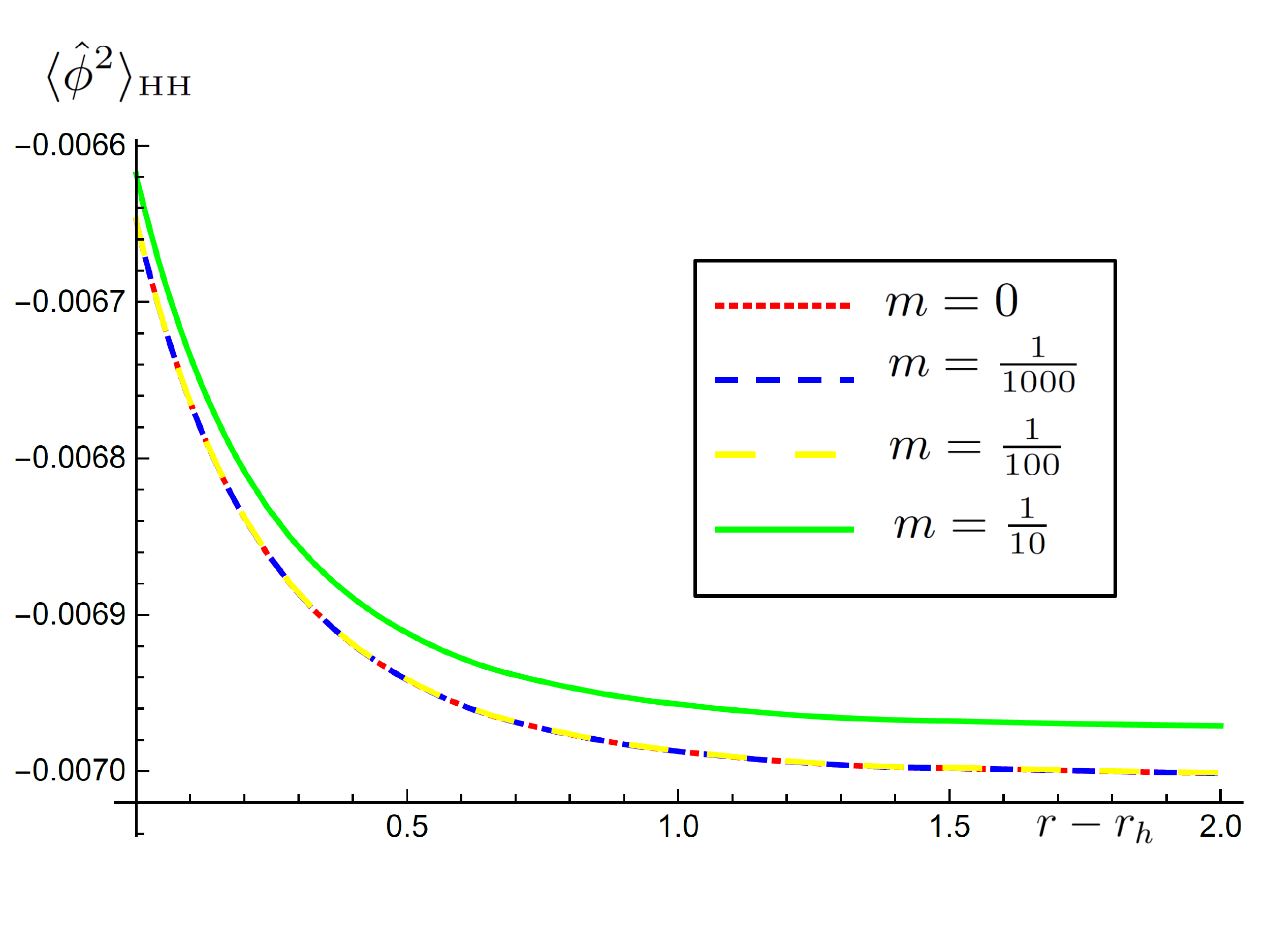}
		\caption{$d=4,\xi=0$}
		\label{fig:d4m}
	\end{subfigure}%
	\begin{subfigure}{.5\textwidth}
		\includegraphics[width=.8\linewidth]{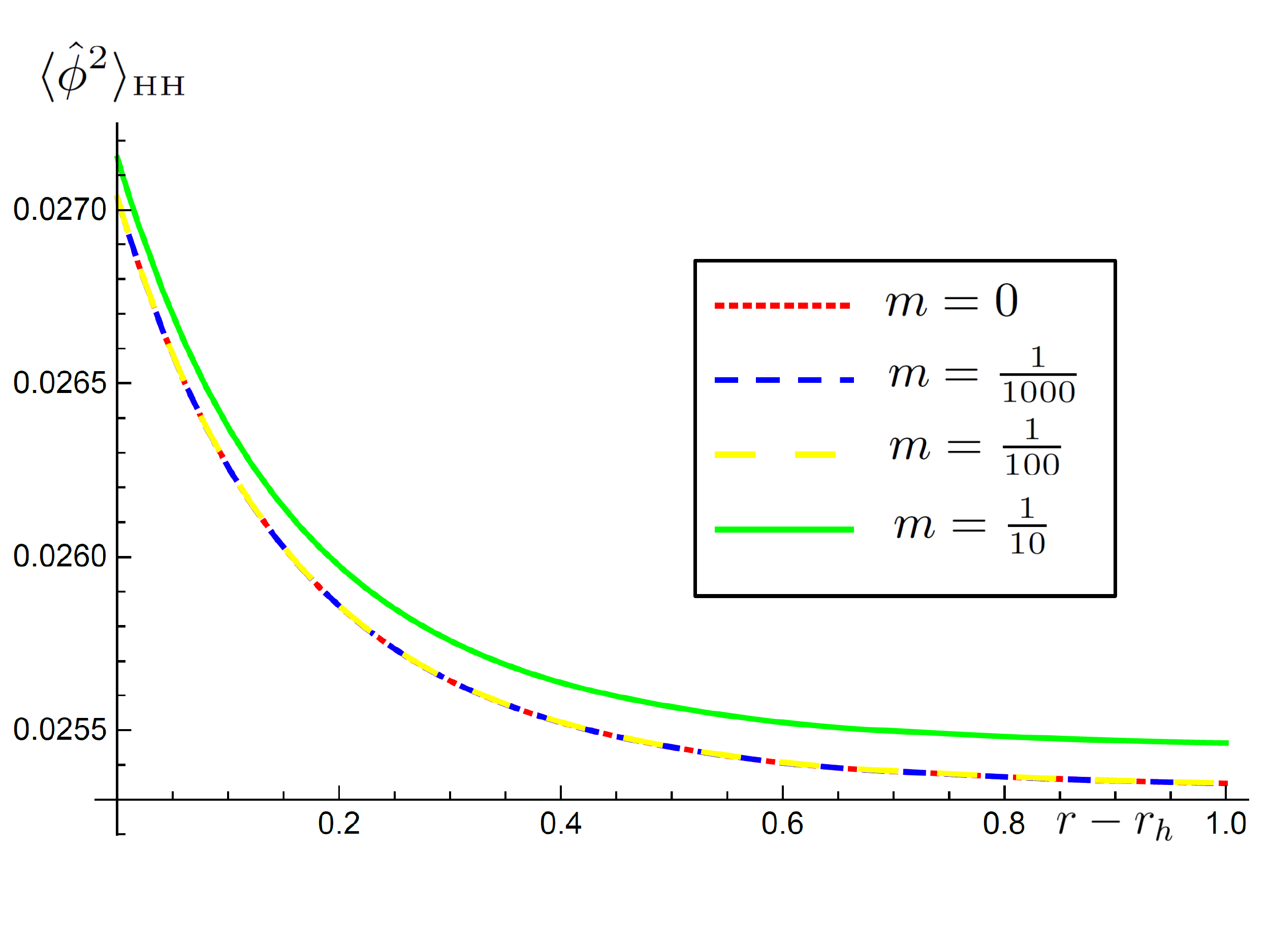}
		\caption{$d=5,\xi=0$}
		\label{fig:d5m}
	\end{subfigure}
	\begin{subfigure}{.5\textwidth}
	\includegraphics[width=.8\linewidth]{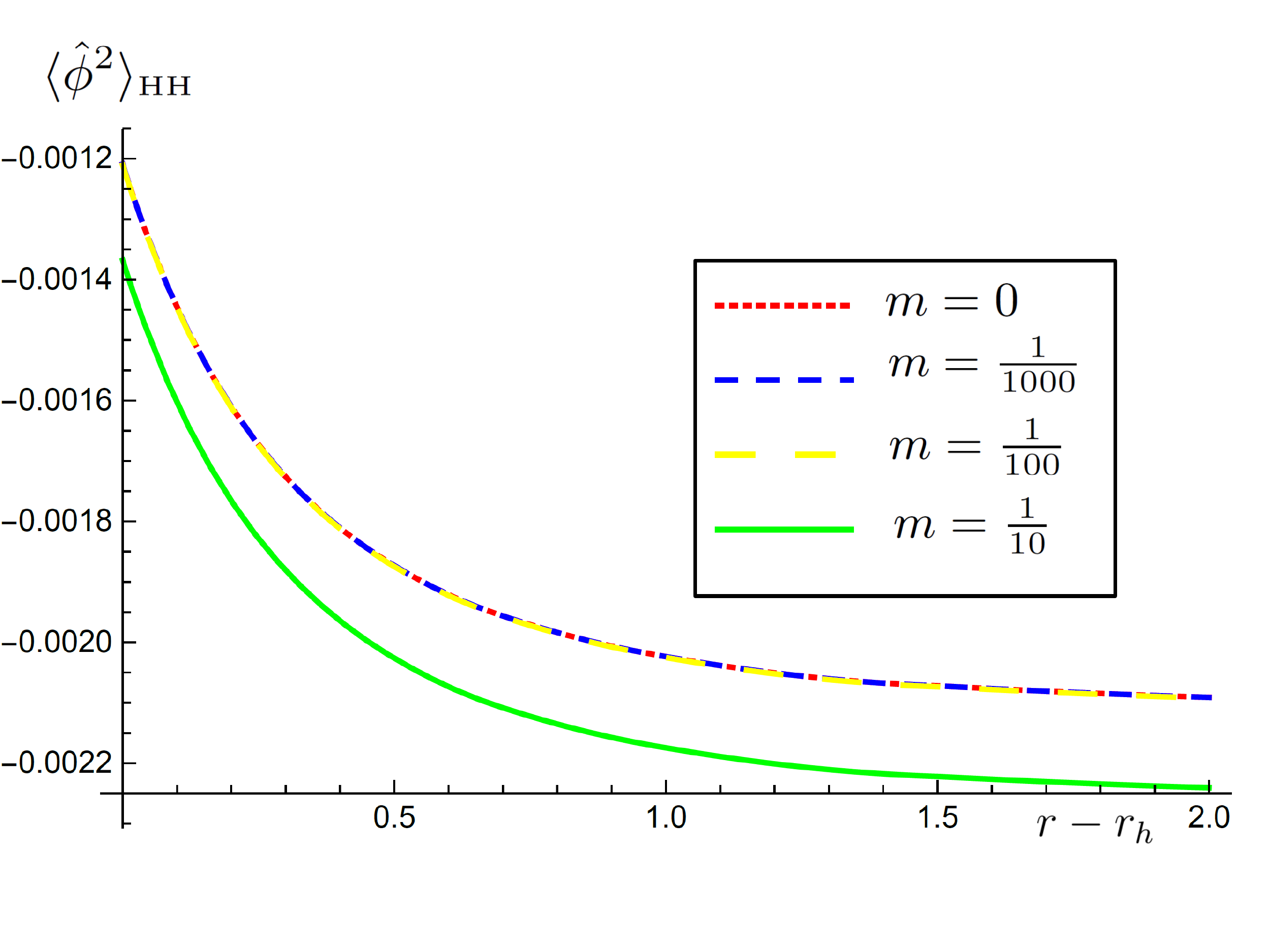}
	\caption{$d=4,\xi=1/6$}
	\label{fig:d4m16}
\end{subfigure}%
	\begin{subfigure}{.5\textwidth}
	\includegraphics[width=.8\linewidth]{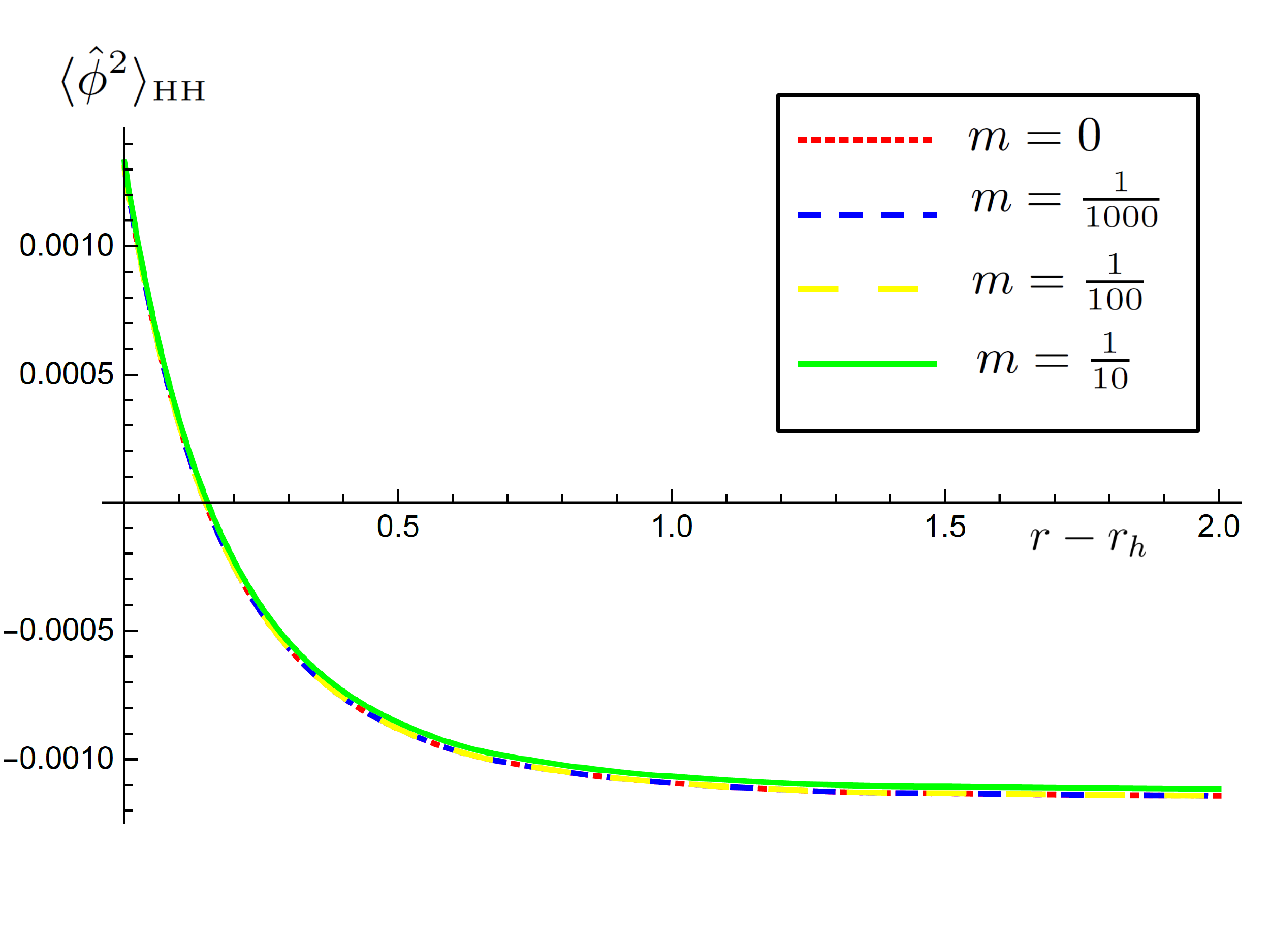}
	\caption{$d=5$}
	\label{fig:d5m16}
\end{subfigure}%
	\caption{\label{fig:m2}Plots of the vacuum polarization as a function of the distance from the black hole for various mass parameters in the range $0\leq m\leq 1/10$. The plots show graphs in this range of mass parameters for $\xi=0$ and $\xi=1/6$ in $d=4$ and $d=5$. The black hole mass parameter $\varpi_{d}$ has been set to twice the adS lengthscale for each $d$.}	
\end{figure*}
The first set of results we discuss are those for which we fix all the parameters except the field mass and examine how the vacuum polarization for a scalar field depends on varying the field mass. We consider field coupling strengths $\xi=0$ and $\xi=1/6$ for spacetime dimensions $d=4$ and $d=5$. We set the mass parameter $\varpi_{d}$ of the black hole to be twice the adS lengthscale, $\varpi_{d}=2\,L=2$. For $d=4$, this corresponds to choosing $M=L=1$. With these choices, we compute the vacuum polarization using the method described above for several values of the field mass. The results are plotted in Fig.~\ref{fig:m1} and Fig.~\ref{fig:m2}. In Fig.~\ref{fig:m1}, we show a 3D plot of the vacuum polarization as a function of field mass $m$ and distance from the black hole; the four distinct plots represent different values of coupling and spacetime dimension. Fig.~\ref{fig:m2} is also comprised of four plots, each of which represent the vacuum polarization as a function of radius only for various values of the field mass. By inspecting these figures, we see that the vacuum polarization is a very slowly varying function of mass near $m=0$, for example, in Fig.~\ref{fig:m2} we see that increasing the mass from $m=0$ to $m=1/100$ makes  no perceptible difference to the value of the vacuum polarization (at all calculated points the difference is of the order of the accuracy our method). This suggests that a massless field approximation is reasonable when the fields present have small but non-zero mass. Moving away from $m\approx 0$, increasing the field mass to $m=1/10$ does significantly affect the vacuum polarization, the vacuum polarization clearly increases compared to the massless case for $\xi=0$ for both $d=4$ and $d=5$ as well as for $\xi=1/6,  d=5$ (although the increase is much less pronounced for this case) and decreases for $\xi=1/6, d=4$. Moreover, the difference between the massive and massless case is not uniform in $r$, rather the difference increases with $r$ until the the graphs asymptote to the $m$-dependent pure adS values. Increasing the field mass even further, we see from Fig.~\ref{fig:m1} that the vacuum polarization becomes a rapidly increasing function of field mass for large mass, which is about $m\gtrsim 1$ (in the $d=4, \xi=1/6$ case it firstly reaches a minimum values in the region of $m=1$ at all radial points).  Indeed this growth appears to be unbounded as $m$ increases. This result is in stark contrast to the analogous result for an asymptotically flat black hole spacetime, where the vacuum polarization decreases as the field mass increases, see for instance \cite{Anderson:1990}.  Given that this large-mass divergence is also a feature of the vacuum polarization for quantum fields in pure adS spacetime \cite{winstanleykent:2014}, we can attribute this behaviour to the asymptotic structure of the geometry. As we will see later, this large-mass behaviour has serious implications for the validity of the DeWitt-Schwinger approximation for the vacuum polarization.


\subsubsection{Varying $\xi$}
		\begin{figure*}
	\begin{subfigure}{.5\textwidth}
		\includegraphics[width=.8\linewidth]{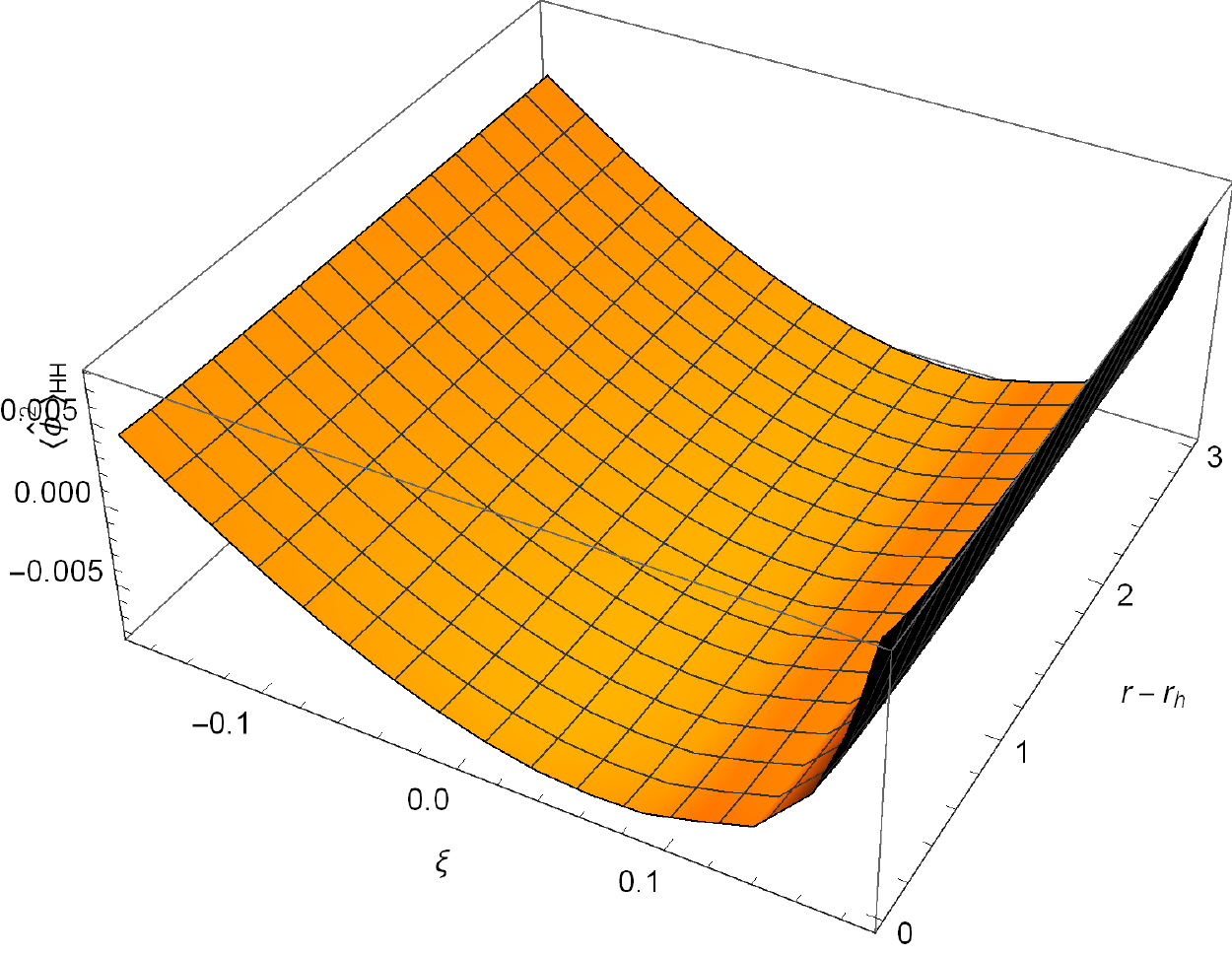}
		\caption{$d=4$}
		\label{fig:d4xi}
	\end{subfigure}%
	\begin{subfigure}{.5\textwidth}
		\includegraphics[width=.8\linewidth]{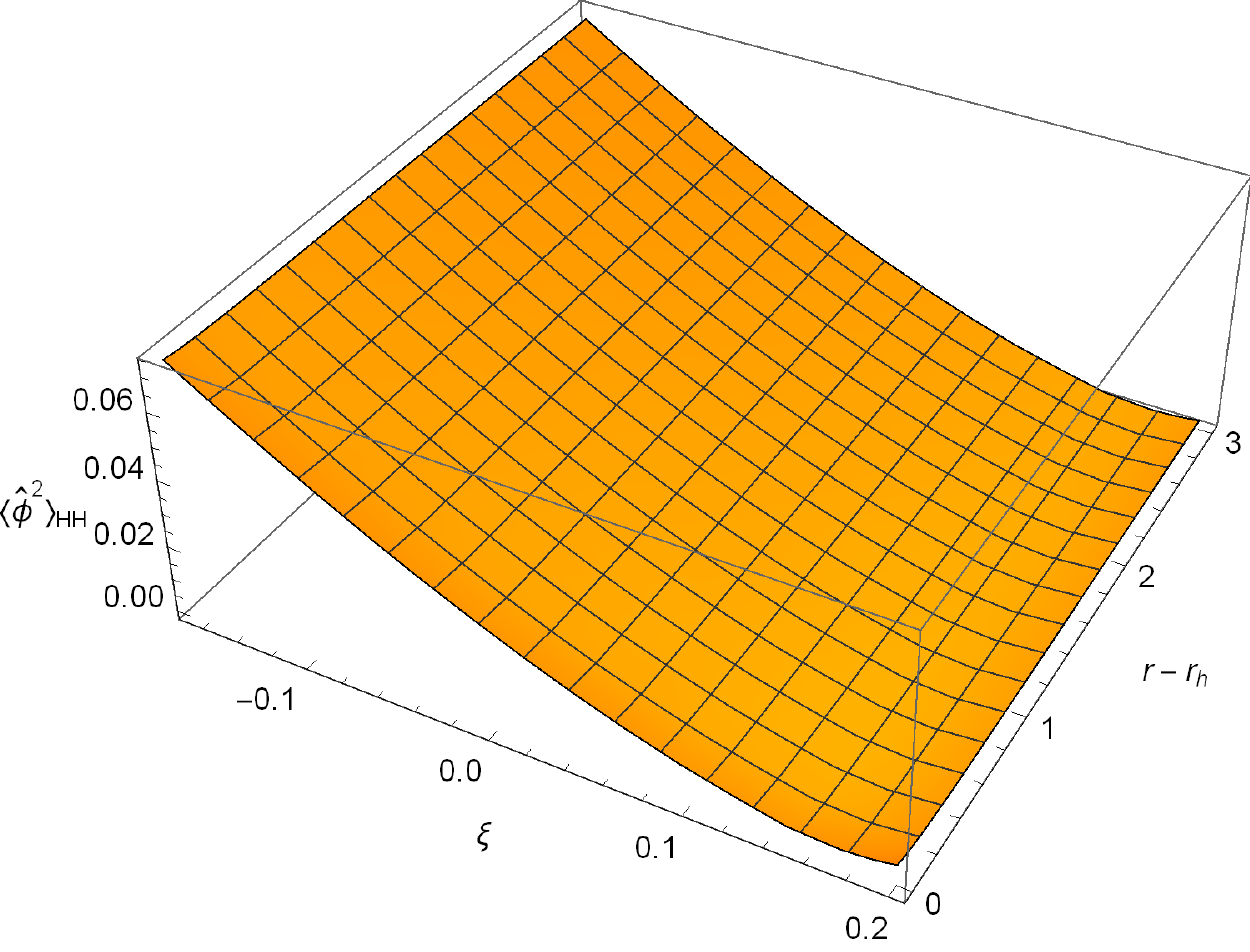}
		\caption{$d=5$}
		\label{fig:d5xi}
	\end{subfigure}
	\begin{subfigure}{.5\textwidth}
		\includegraphics[width=.8\linewidth]{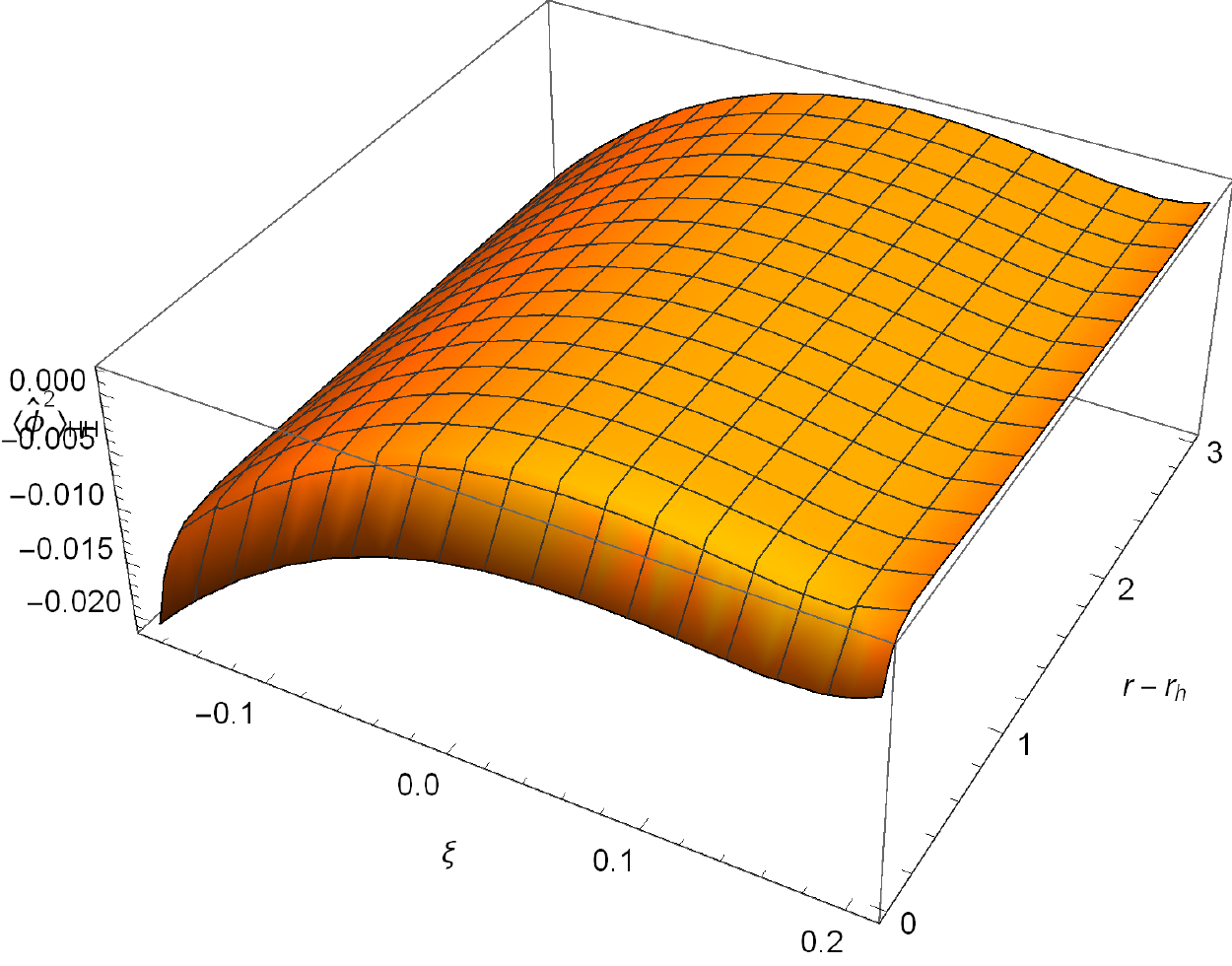}
		\caption{$d=6$}
		\label{fig:d6xi}
	\end{subfigure}%
	\begin{subfigure}{.5\textwidth}
		\includegraphics[width=.8\linewidth]{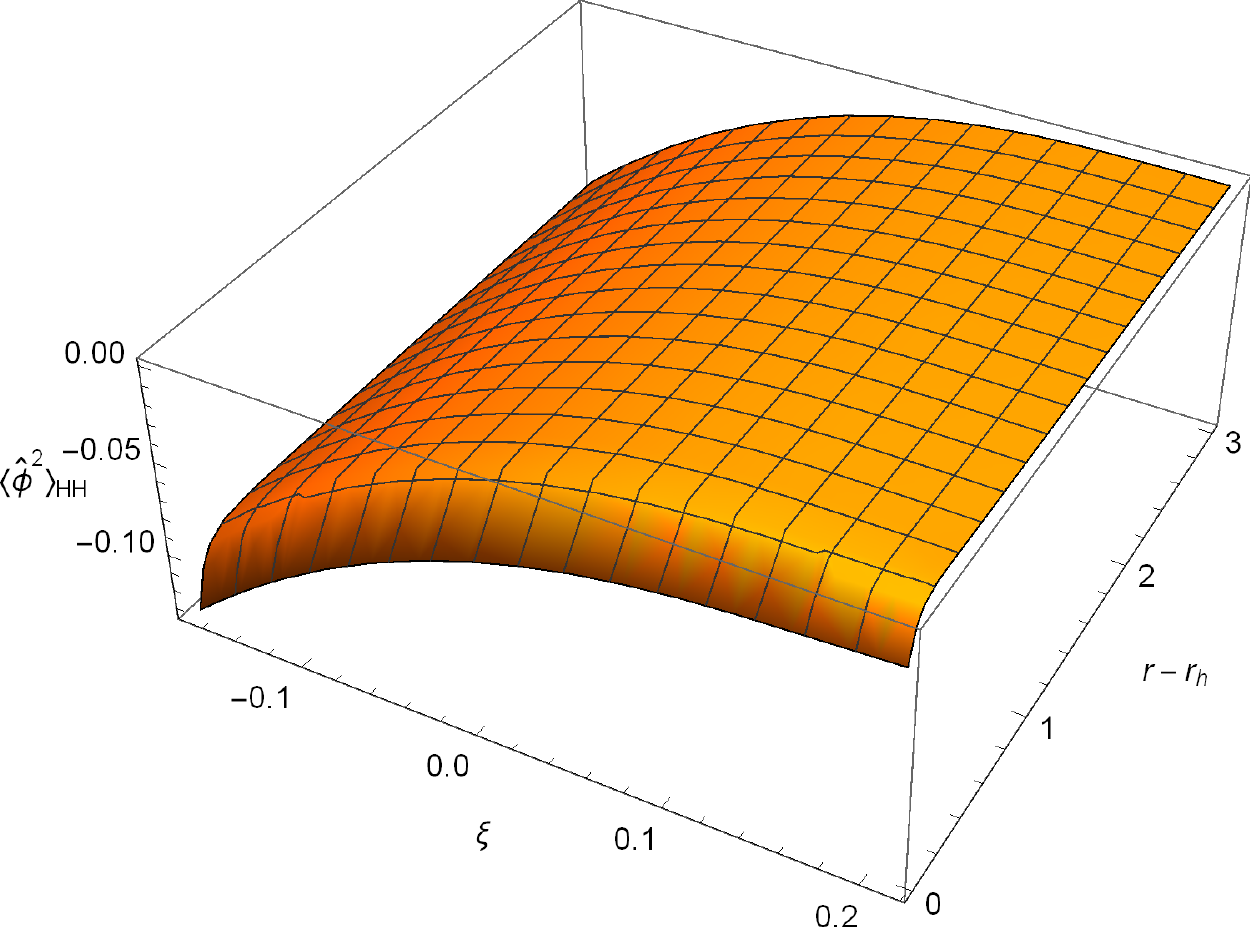}
		\caption{$d=7$}
		\label{fig:d7ofxi}
	\end{subfigure}
	\caption{\label{fig:xi1}Plots of the vacuum polarization for a massless scalar field in SadS spacetime for $d=4$, $d=5$, $d=6$ and $d=7$ as a function of the coupling constant $\xi$ and distance from the black hole. The black hole mass parameter has been set to $\varpi_{d}=2$ for each $d$.}	
\end{figure*}
\begin{figure*}
\begin{subfigure}{.5\textwidth}
	\includegraphics[width=.8\linewidth]{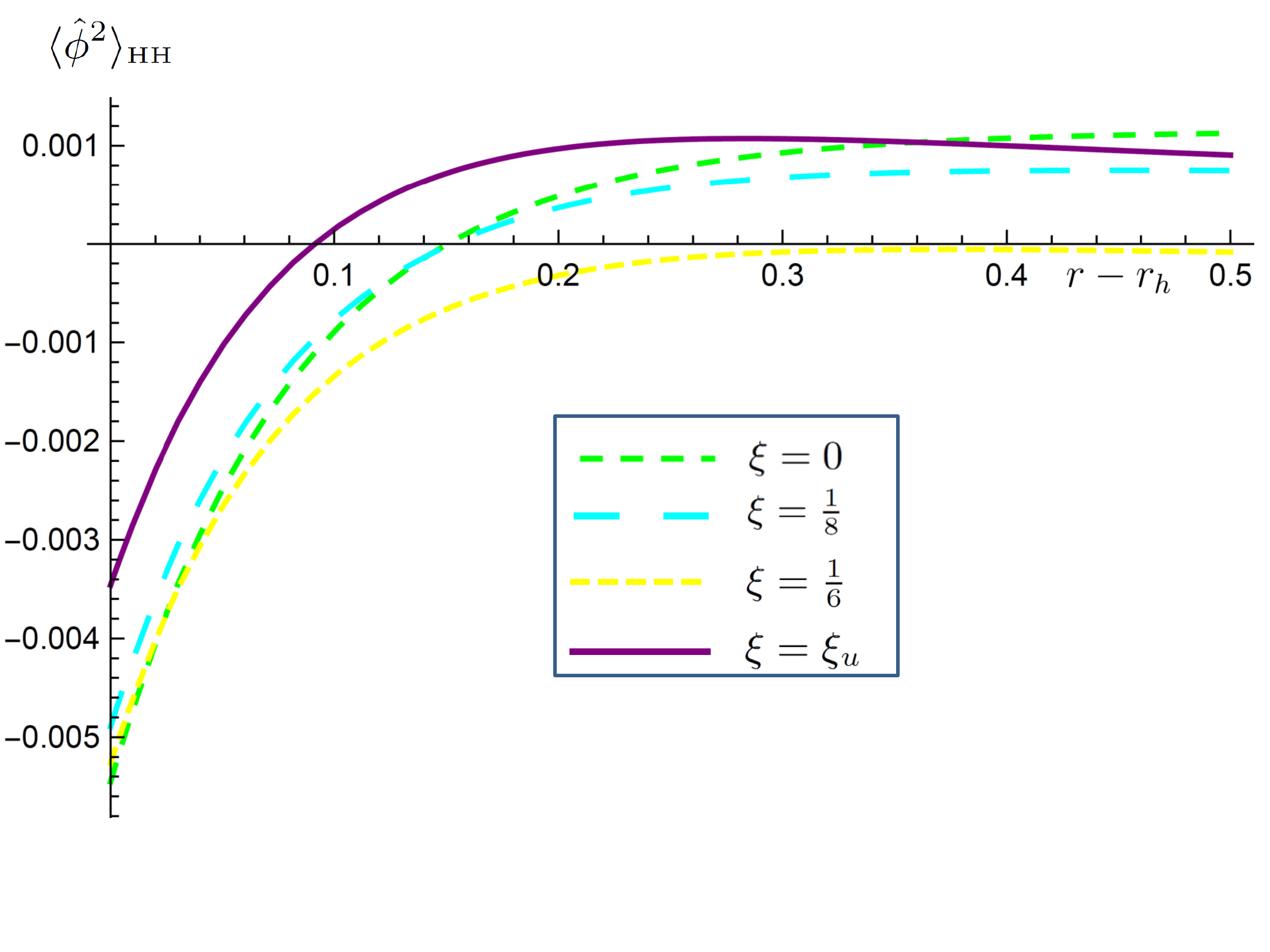}
	\caption{$d=6$ for positive values of $\xi$}
	\label{fig:d6xizoom}
\end{subfigure}%
\begin{subfigure}{.5\textwidth}
	\includegraphics[width=.8\linewidth]{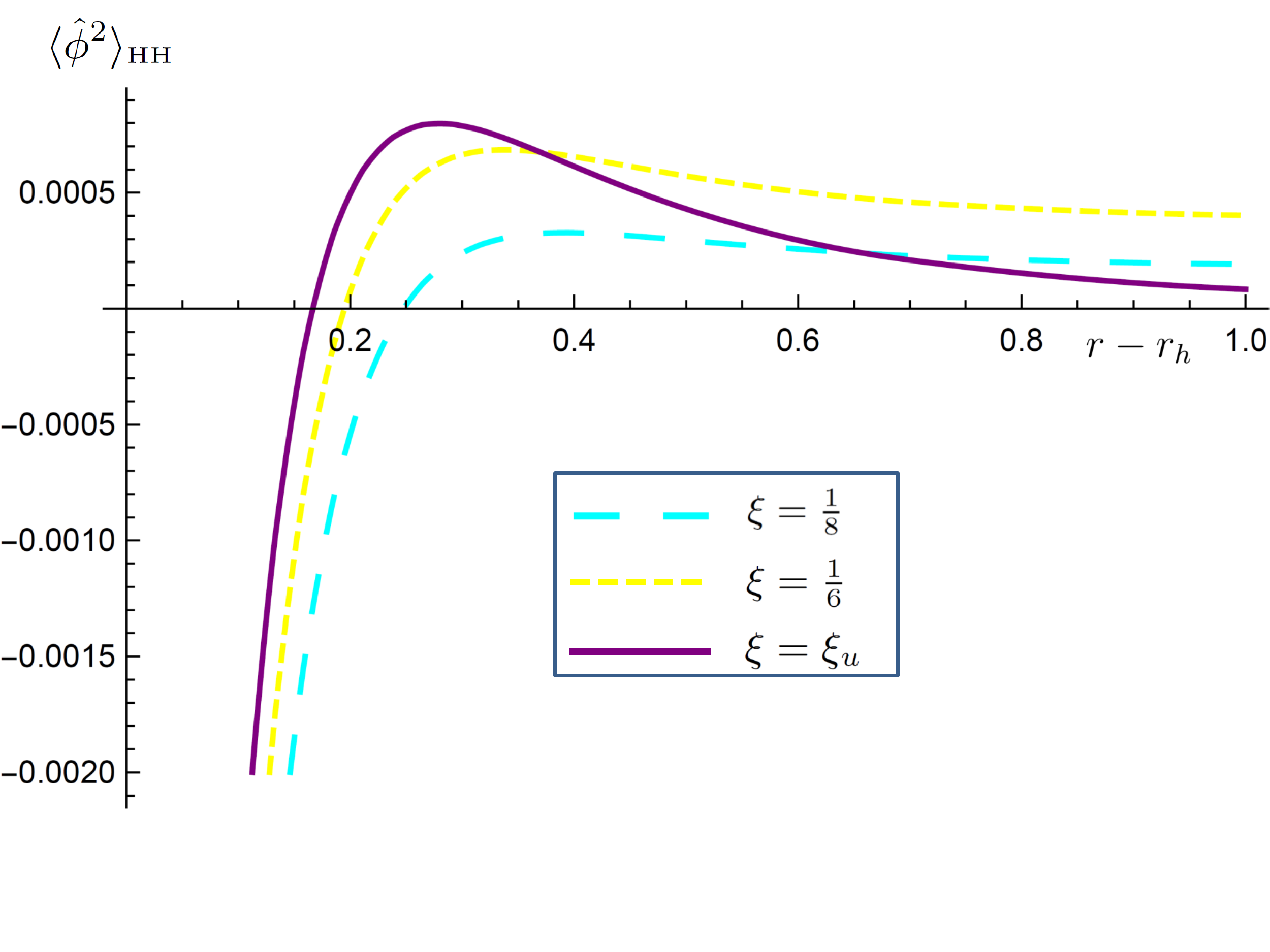}
	\caption{$d=7$ for $\xi \geq 1/8$}
	\label{fig:d7xizoom}
\end{subfigure}
	\caption{\label{fig:xi2}Plots of the vacuum polarization as a function of distance from the black hole for a massless scalar field in SadS spacetime for $d=6$ and $d=7$ with different values of the coupling constant $\xi$. The black hole mass parameter is  $\varpi_{d}=2$ in units where $L=1$. }	
\end{figure*}
The next set of results we present are those for which the parameters of the background geometry are fixed as before, the field is massless, and we examine the dependence of the vacuum polarization on the coupling of the field to the background curvature. These results are presented in Fig.~\ref{fig:xi1} and Fig.~\ref{fig:xi2}. In Fig.~\ref{fig:xi1}, we show 3D plots of the vacuum polarization as a function of the coupling and the distance from the black hole for $d=4$, $d=5$ $d=6$ and $d=7$ In Fig.~\ref{fig:xi2}, we also present additional plots of the vacuum polarization as a function of the distance from the black hole for various values of $\xi$ in the $d=6$ and $d=7$ cases.

In the $d=4$ case, we see from Fig.~\ref{fig:xi1} that  $\langle \hat{\phi}^{2} \rangle_{\subHH}$ has a minimum for a small positive $\xi$-value. For $d=5$, we see $\langle \hat{\phi}^{2} \rangle_{\subHH}$ again has a minimum value, this time it occurs in the region of $\xi=1/6$. $\langle \hat{\phi}^{2} \rangle_{\subHH}$ then increases slightly as $\xi \to \xi_u=1/5$. We note that for the coupling's maximum allowed value $\xi_u$, $\langle \hat{\phi}^{2} \rangle_{\subHH} \to 0$ as $r \to \infty$, as is the case for a massless field in odd dimensional pure adS \cite{winstanleykent:2014}.

 For $d=6$, the dependence of $\langle \hat{\phi}^{2} \rangle_{\subHH}$ has two turning points in $\xi$. The vacuum polarization firstly increases with increasing $\xi$, reaching a local $r$-dependent maximum (this maximum occurs in the region of $\xi=1/8$ near the black hole or $\xi=0$ further away from the horizon, see Fig.~\ref{fig:d6xizoom}). The vacuum polarization then decreases to a local minimum value in the region of $\xi=1/6$ (the turning point here appears to be approximately uniform in $r$) before increasing again as $\xi \to \xi_u=5/24$. We also see that the maximum value of $\langle \hat{\phi}^{2} \rangle_{\subHH}$ occurs for $\xi=\xi_u$ near the black hole and in the vicinity of $\xi=0$ further away.

In the $d=7$ case, in the region of the black hole horizon, $\langle \hat{\phi}^{2} \rangle_{\subHH}$ increases monotonically with increasing $\xi$ until it reaches its maximum value at $\xi=\xi_u=3/14$. The dependence on $\xi$  differs further from the horizon however, though this is not easy to see from the 3D plot, in this region $\langle \hat{\phi}^{2} \rangle_{\subHH}$ reaches a maximum value in the region of $\xi=1/6$ before decreasing to its value at $\xi=\xi_u$, which again approaches 0 as $r \to \infty$, see Fig.~\ref{fig:d7xizoom}.

Interestingly, in all cases  the vacuum polarization can become arbitrarily large (in the positive direction for $d=4,5$ and the negative direction for $d=6,7$) by taking the coupling constant to be increasingly negative, without any obvious violations of the semi-classical approximation. This mirrors the divergence of  $\langle \hat{\phi}^{2} \rangle_{\subHH}$ for large field mass. This is not surprising since examination of Eq.~(\ref{eq:radialeqn}) implies that the Green function depends on $m$ and $\xi$ only through the effective mass $\mu_{\xi}$, and this effective mass is degenerate with distinct pairs of values for $m$ and $\xi$. For example, a massive minimally coupled field has the same effective mass as a massless non-minimally coupled field so long as $m^{2}=-d(d-3)\xi/L^{2}$, where $m$ is the mass of the massive field and $\xi$ the coupling strength of the non-minimally coupled massless field. We note that this implies that the vacuum polarization should get increasingly negative for large values of the field mass for $d=6$, $d=7$. Though we have not presented any plots for large field mass in these dimensions, we have checked that the vacuum polarization does decrease without bound for increasing mass, as expected. 
 
 That the vacuum polarization can become arbitrarily large without violating any obvious assumption in the semi-classical approximation is troubling since it could ostensibly lead to large back-reaction effects on the classical geometry. Had we chosen different boundary conditions other than the Dirichlet ones, we could have ruled out arbitrarily negative couplings by appealing to the fact that well-posedness of the Klein Gordon equation requires
 \begin{align}
     -\frac{1}{4}<\mu_{\xi}<\frac{3}{4},
 \end{align}
 for Neumann and Robin boundary conditions \cite{Warnick:2012fi}. Translating this as an inequality for $\xi$ implies
 \begin{align}
     \frac{L^{2}m^{2}}{d(d-1)}+\frac{(d-3)(d+1)}{4 d (d-1)}<\xi<\frac{L^{2}m^{2}}{d(d-1)}+\frac{d-1}{4 d}.
 \end{align}
 However, in our case, well-posedness for Dirichlet boundary conditions is guaranteed so long as the Breitenlohner-Freedman (\ref{eq:ximax}) bound is satisfied \cite{Holzegel:2011qj, Vasy} and there is no obvious way to rule out arbitrarily negative values of the coupling leading to potentially arbitrarily large back-reaction. Comparing the differences in the stress-energy tensors and the back-reaction on the SadS background for the field satisfying different boundary conditions may offer some insights into this problem.

\subsubsection{Varying $M$}
	\begin{figure*}
		\begin{subfigure}{.5\textwidth}
			\includegraphics[width=.8\linewidth]{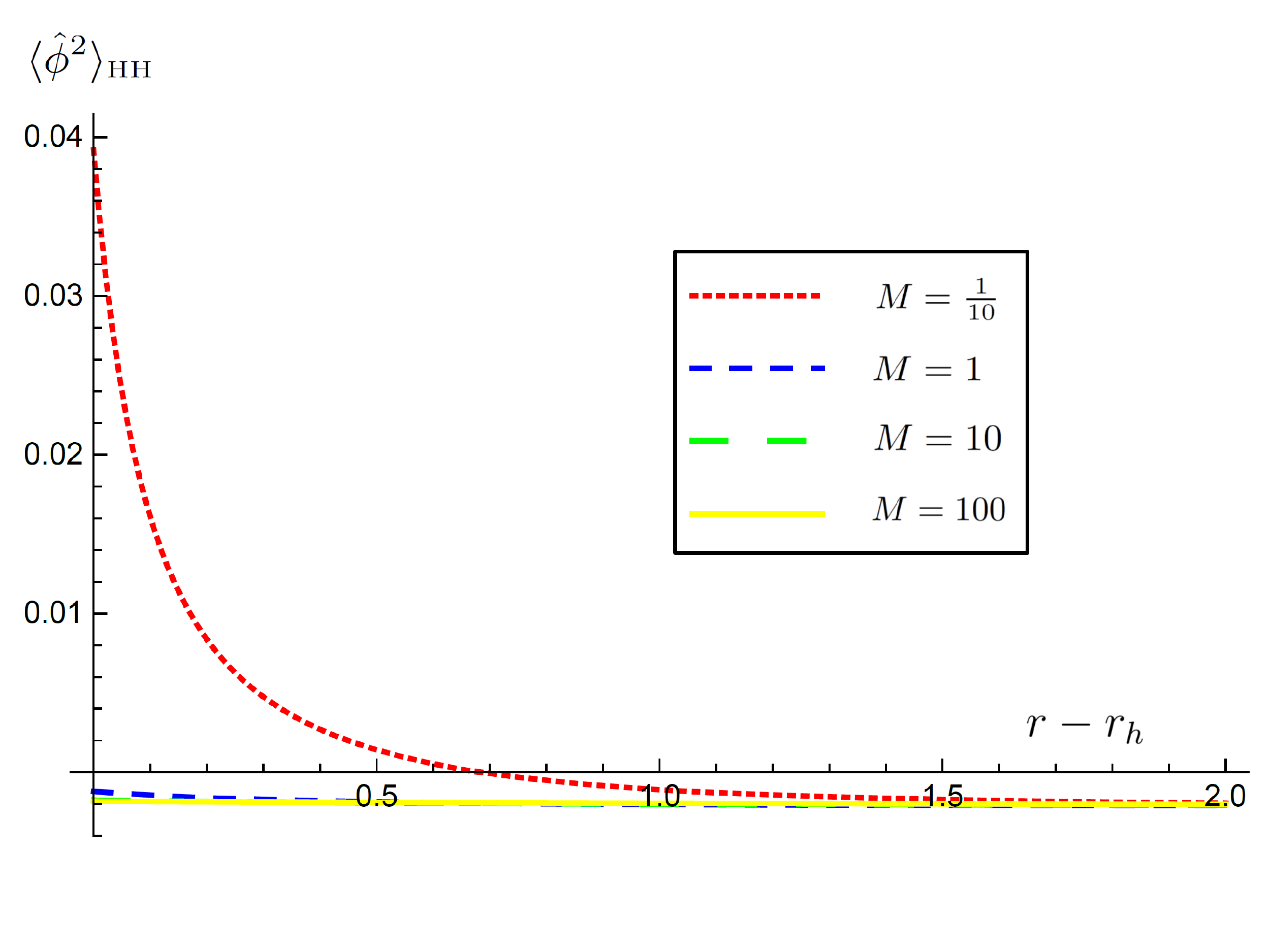}
			\caption{$d=4$}
			\label{fig:d4M}
		\end{subfigure}%
		\begin{subfigure}{.5\textwidth}
			\includegraphics[width=.8\linewidth]{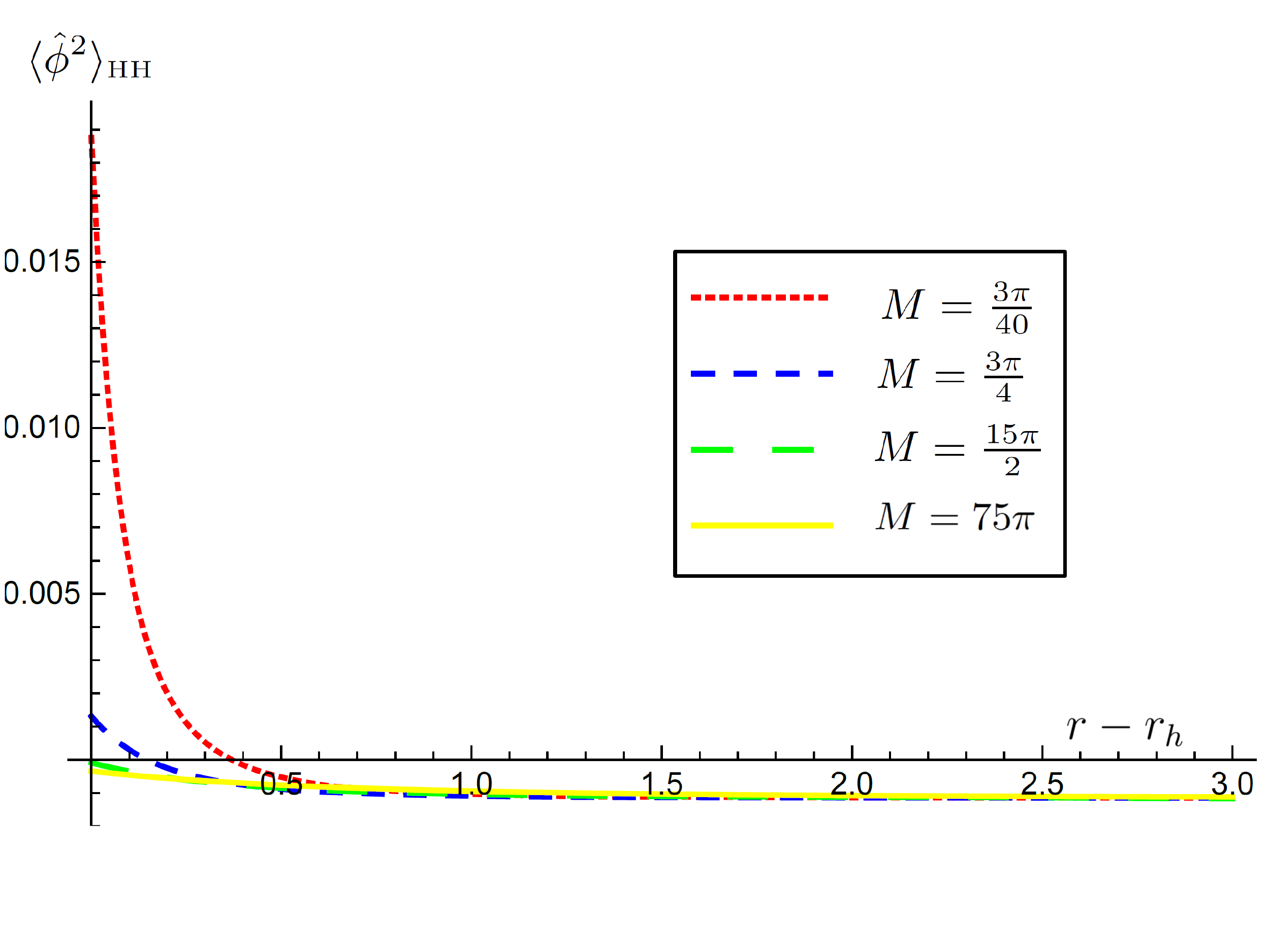}
			\caption{$d=5$}
			\label{fig:d5M}
		\end{subfigure}
		\caption{\label{fig:M} Plots of the vacuum polarization for a massless scalar field in SadS spacetime for $d=4$ and $d=5$ with different values of the black hole mass parameter $\varpi_{d}$. The scalar coupling constant is $\xi=1/6$}	
	\end{figure*}

	\begin{figure*}
		\begin{subfigure}{.5\textwidth}
			\includegraphics[width=.8\linewidth]{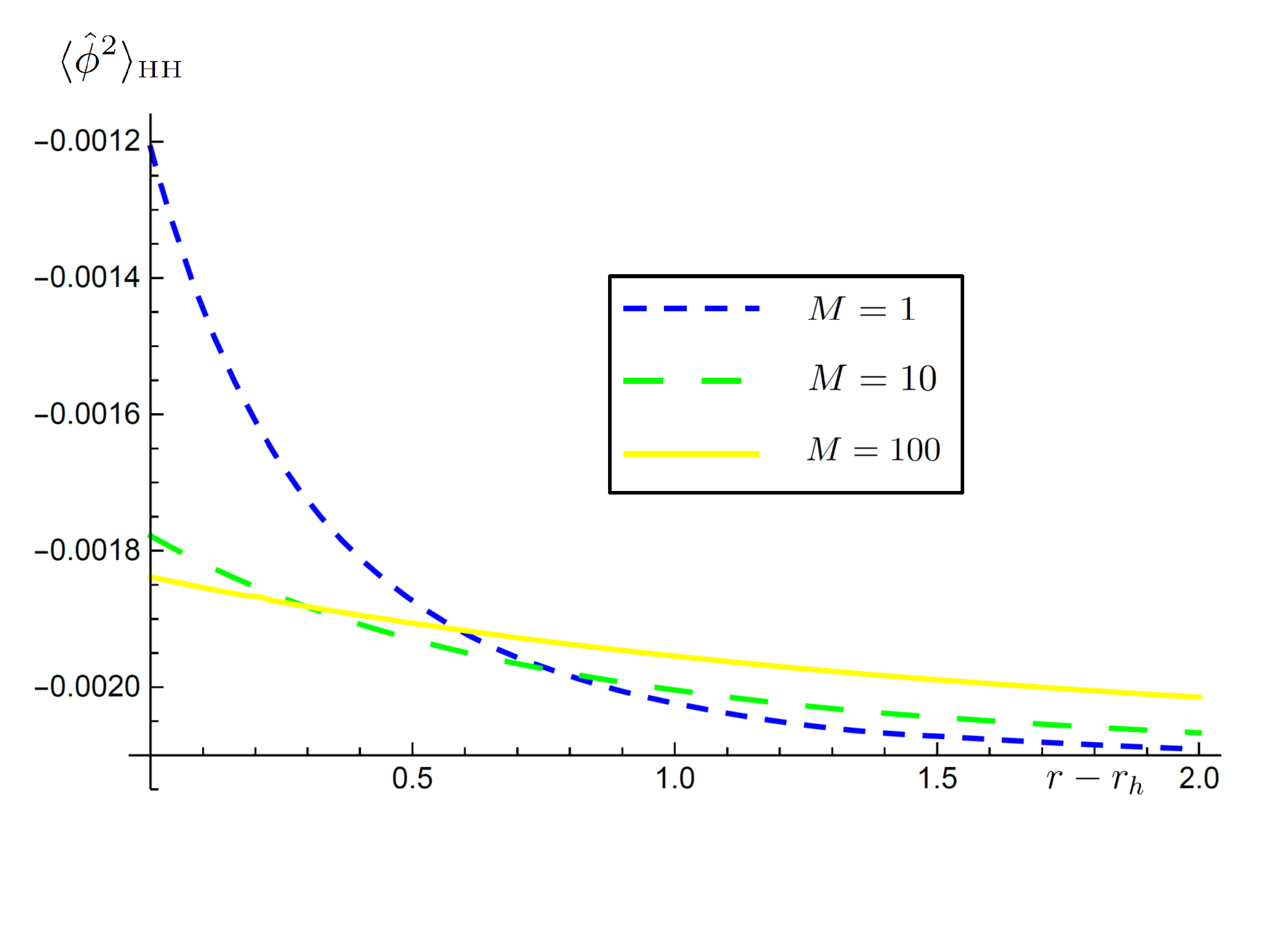}
			\caption{$d=4$}
			\label{fig:d4Mlarge}
		\end{subfigure}%
		\begin{subfigure}{.5\textwidth}
			\includegraphics[width=.8\linewidth]{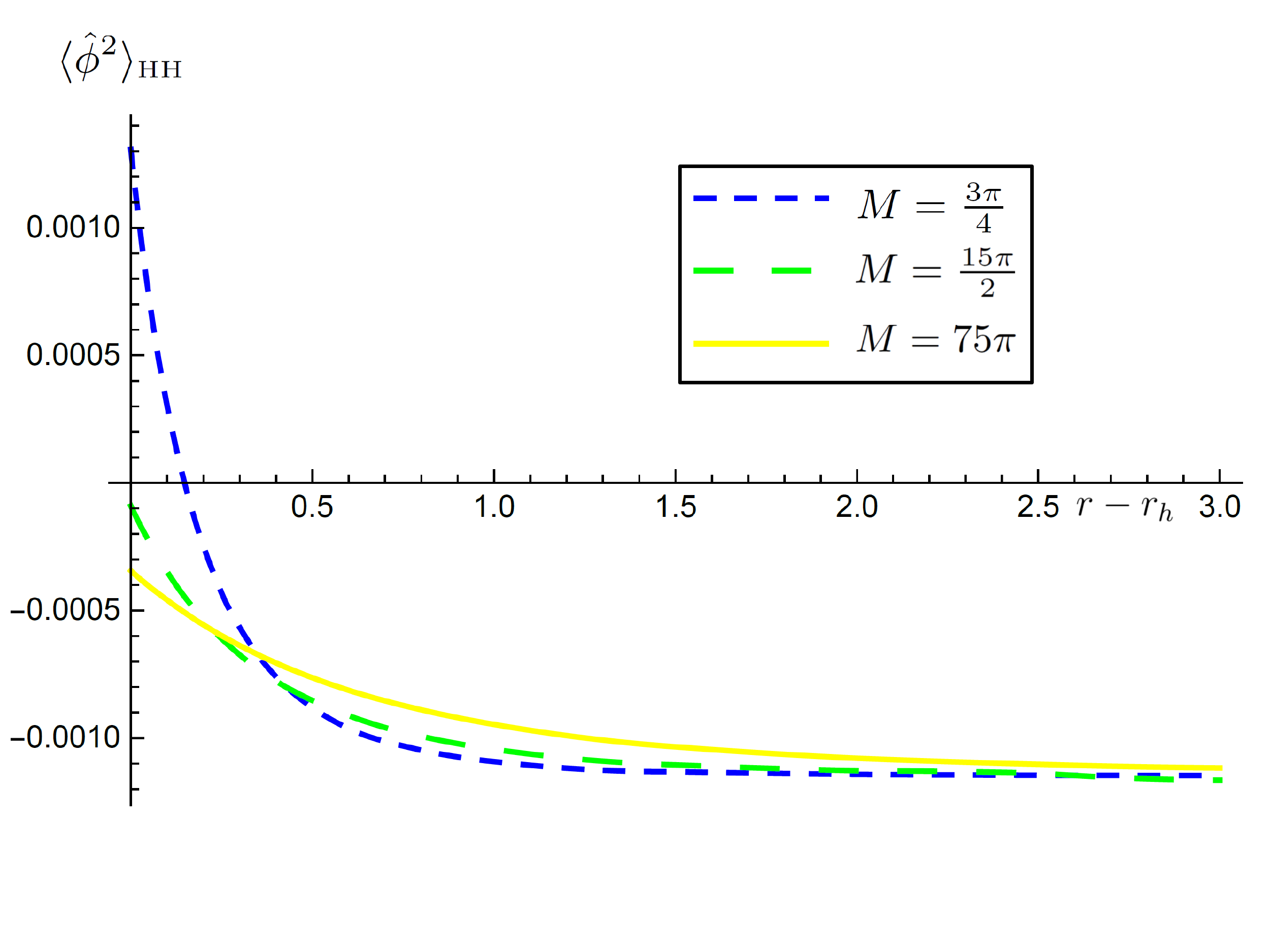}
			\caption{$d=5$}
			\label{fig:d5large}
		\end{subfigure}
		\caption{\label{fig:Mlarge}Plots of the vacuum polarization for a massless scalar field in SadS spacetime for $d=4$ and $d=5$ with  values of the black hole mass parameter  $\varpi_{d}=2,20,200$. The coupling is $\xi=1/6$.}	
	\end{figure*}
In this subsection, we present plots for the vacuum polarization keeping the field parameters fixed and varying the black hole mass. It is easier to specify the mass in terms of the parameter $\varpi_{d}$ rather than $M$; they are related by (\ref{eq:mass}). In particular,  Fig.~\ref{fig:M} shows plots of $\langle \hat{\phi}^{2} \rangle_{\subHH}$ for a massless conformally-coupled field for both $d=4$ and $d=5$, with black hole mass parameters $\varpi_{d}=1/5,\, 2,\, 20,\, 200$. For $d=4$, these correspond to $M=1/10,\,1,\,10,\,100$, respectively. For $d=5$, these values correspond to black hole mass $M=3\pi/40,\,3\pi/4,\,15\pi/2,\,75\pi$, respectively. Since the graph for $\varpi_{d}=1/5$ dominates the other graphs, in Fig.~\ref{fig:Mlarge} we plot $\langle \hat{\phi}^{2} \rangle_{\subHH}$ for the other mass parameters, excluding $\varpi_{d}=1/5$.  Recall that this is in units where the adS lengthscale is $L=1$. We firstly note that the asymptotic adS value is independent of $M$ and so $\langle \hat{\phi}^{2} \rangle_{\subHH}$ ought to approach the same asymptotic value as $r \to \infty$ for each $M$ being considered. We can see from Fig.~\ref{fig:M} that this is indeed the case and that this convergence occurs at a faster rate in $d=5$ than in $d=4$, as one would expect from the form of the metric function (\ref{eq:fSadS}). In both the $d=4$ and $d=5$, the vacuum polarization is a decreasing function of mass in the vicinity of the black hole horizon, while further away the dependence is more complicated. For example, in the plots shown in Fig.~\ref{fig:Mlarge}, from around $r=r_{\textrm{h}}+0.5$ where the graphs approximately intersect, the vacuum polarization appears to be an increasing function of $M$, possibly changing again even further from the horizon as suggested by the intersection of the $d=5$ graphs in Fig.~\ref{fig:Mlarge} at about $r=r_{\textrm{h}}+3$.  

There is a marked difference in the functional dependence of the temperature on the mass in the SadS black hole compared to the Schwarzschild black hole, the latter being a decreasing function of mass. In the SadS black hole spacetime, however, the black hole temperature (as a function of $M$ for fixed $L$) has a minimum. For $d=4$, this occurs at $M_{\textrm{min}}=2 L/(3\sqrt{3})$, or equivalently $\varpi_{4}^{\textrm{min}}=4 L/(3\sqrt{3})$, while for $d=5$ we have $M_{\textrm{min}}=9\pi L^{2}/32$, or equivalently $\varpi_{5}^{\textrm{min}}=3 L^{2}/4$. Moreover, the temperature is a rapidly decreasing function near $M=0$, and a very slowly increasing function of mass for $M>M_{\textrm{min}}$. Since the temperature is due to the vacuum polarization, this dependence of the temperature on the mass ought to be reflected in the vacuum polarization plots for varying mass. This can indeed be seen by the fact that the vacuum polarization for distinct $M>M_{\textrm{min}}$ are nearly indistinguishable from each other (and from the constant adS value) except very close to the horizon, while for small $M$ there is a pronounced increase in the vacuum polarization as expected.
\subsubsection{Varying $d$}
\begin{figure*}
	\begin{subfigure}{.5\textwidth}
		\includegraphics[width=.8\linewidth]{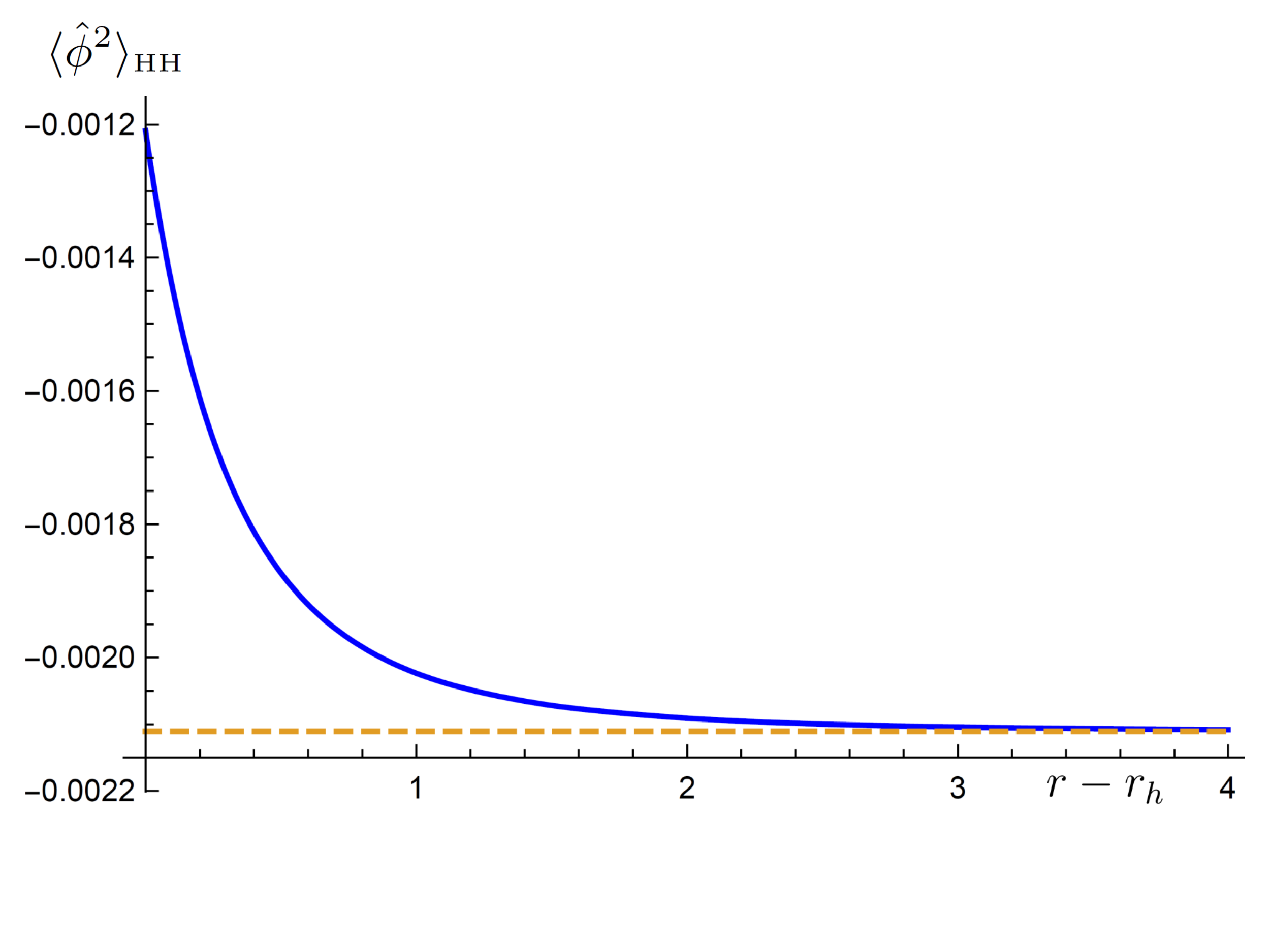}
		\caption{$d=4$}
		\label{fig:d4}
	\end{subfigure}%
	\begin{subfigure}{.5\textwidth}
		\includegraphics[width=.8\linewidth]{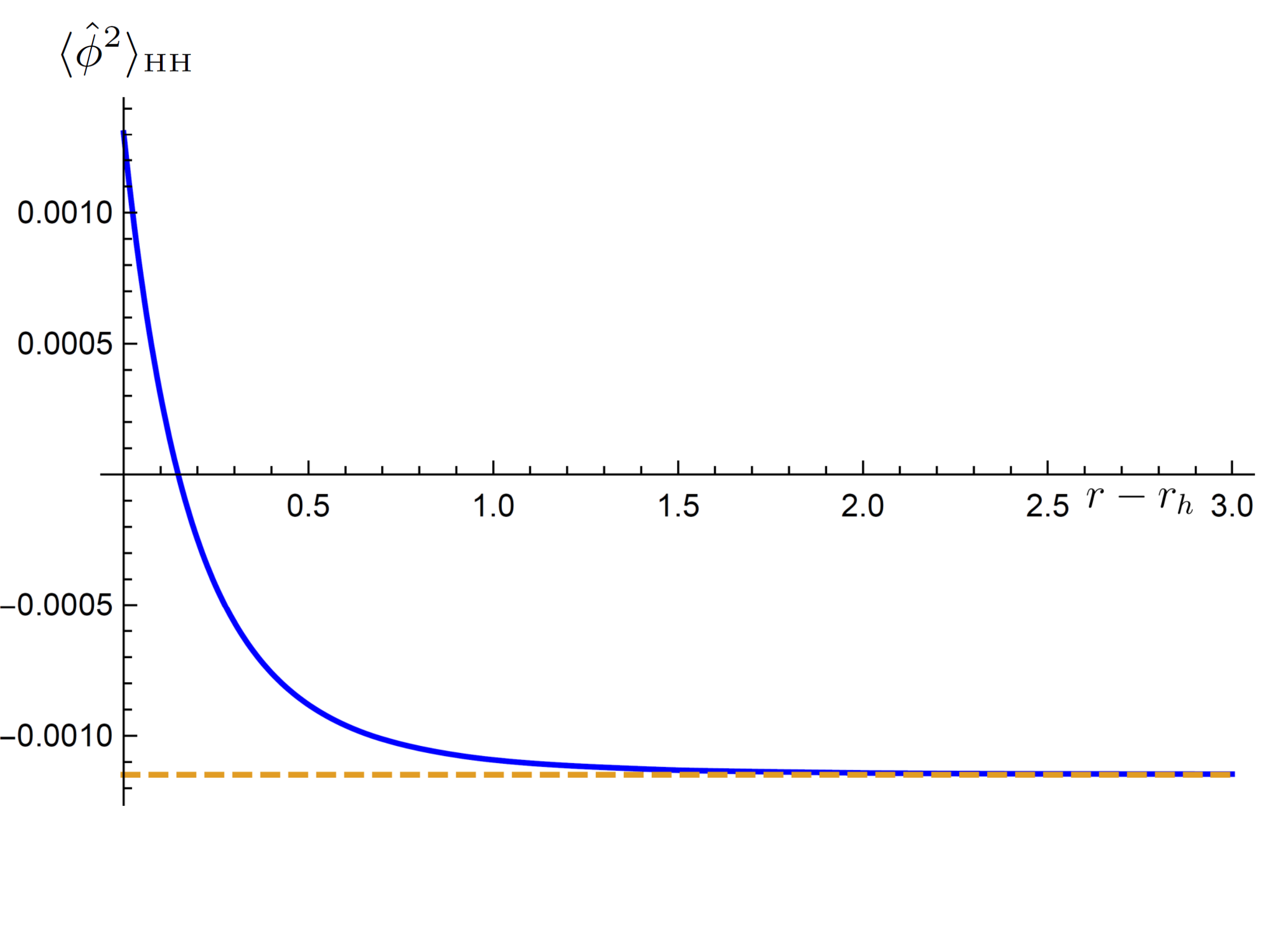}
		\caption{$d=5$}
		\label{fig:d5}
	\end{subfigure}
	\begin{subfigure}{.5\textwidth}
		\includegraphics[width=.8\linewidth]{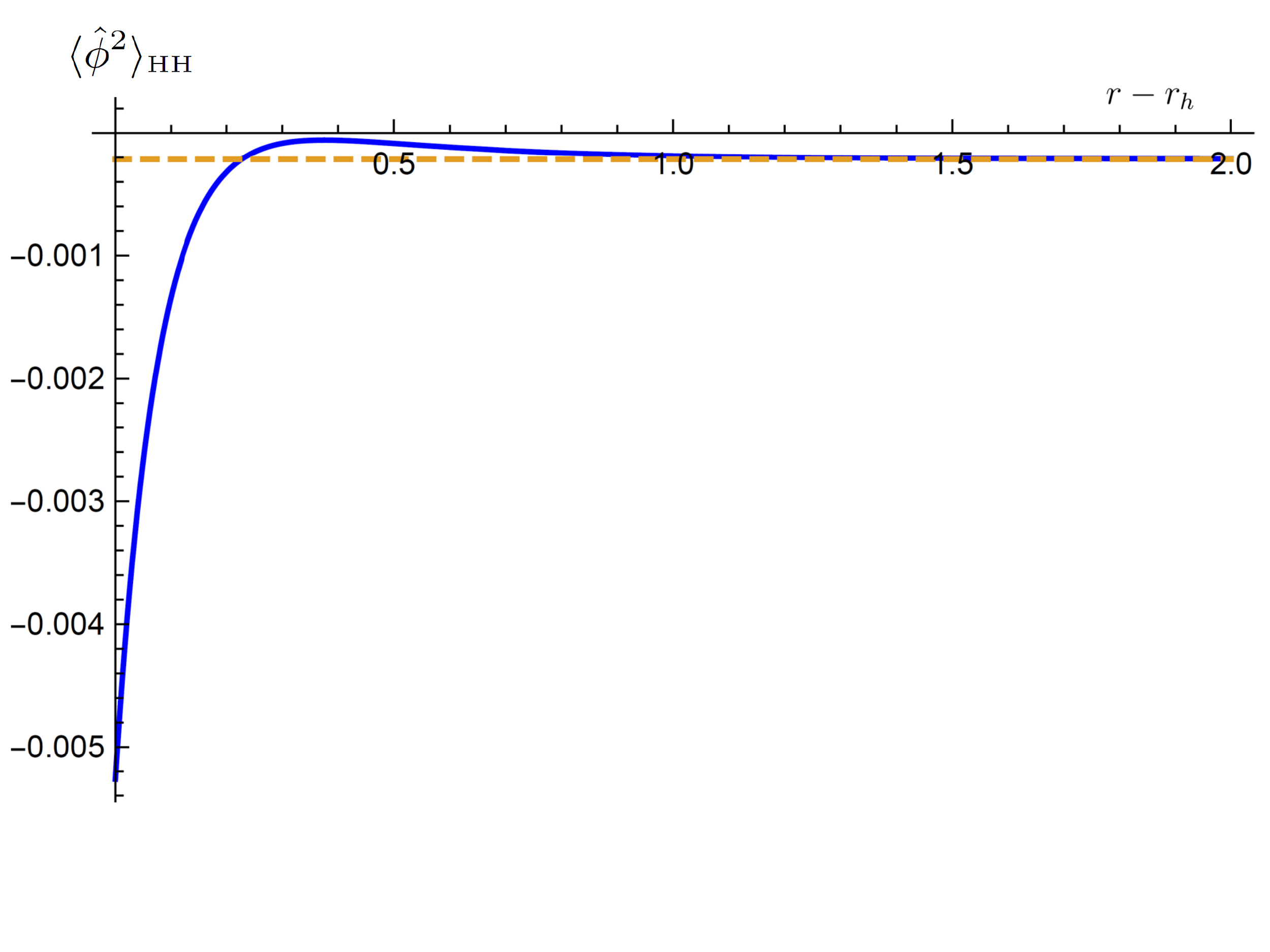}
		\caption{$d=6$}
		\label{fig:d6}
	\end{subfigure}%
	\begin{subfigure}{.5\textwidth}
		\includegraphics[width=.8\linewidth]{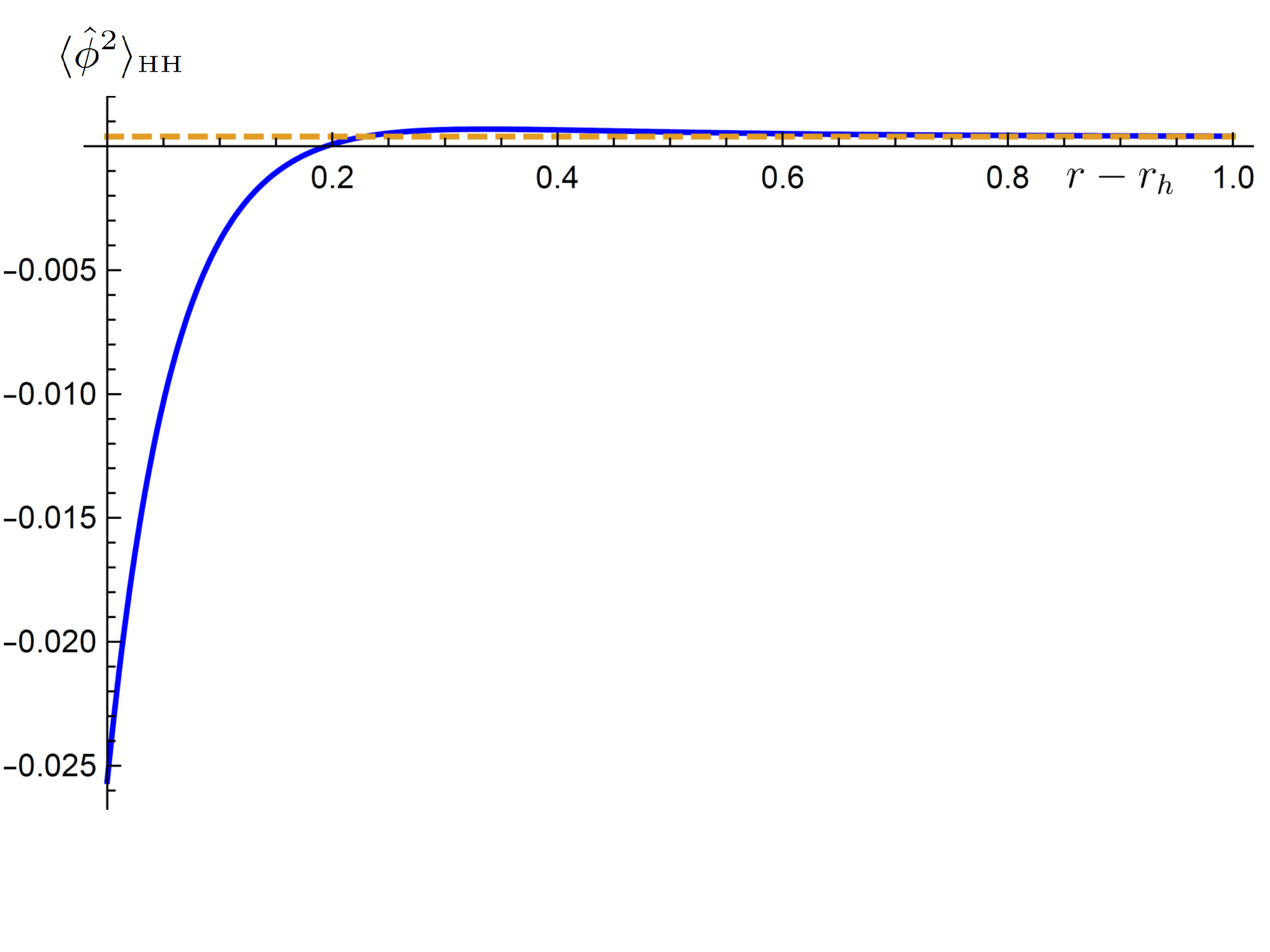}
		\caption{$d=7$}
		\label{fig:d7}
	\end{subfigure}
	\begin{subfigure}{.5\textwidth}
		\includegraphics[width=.8\linewidth]{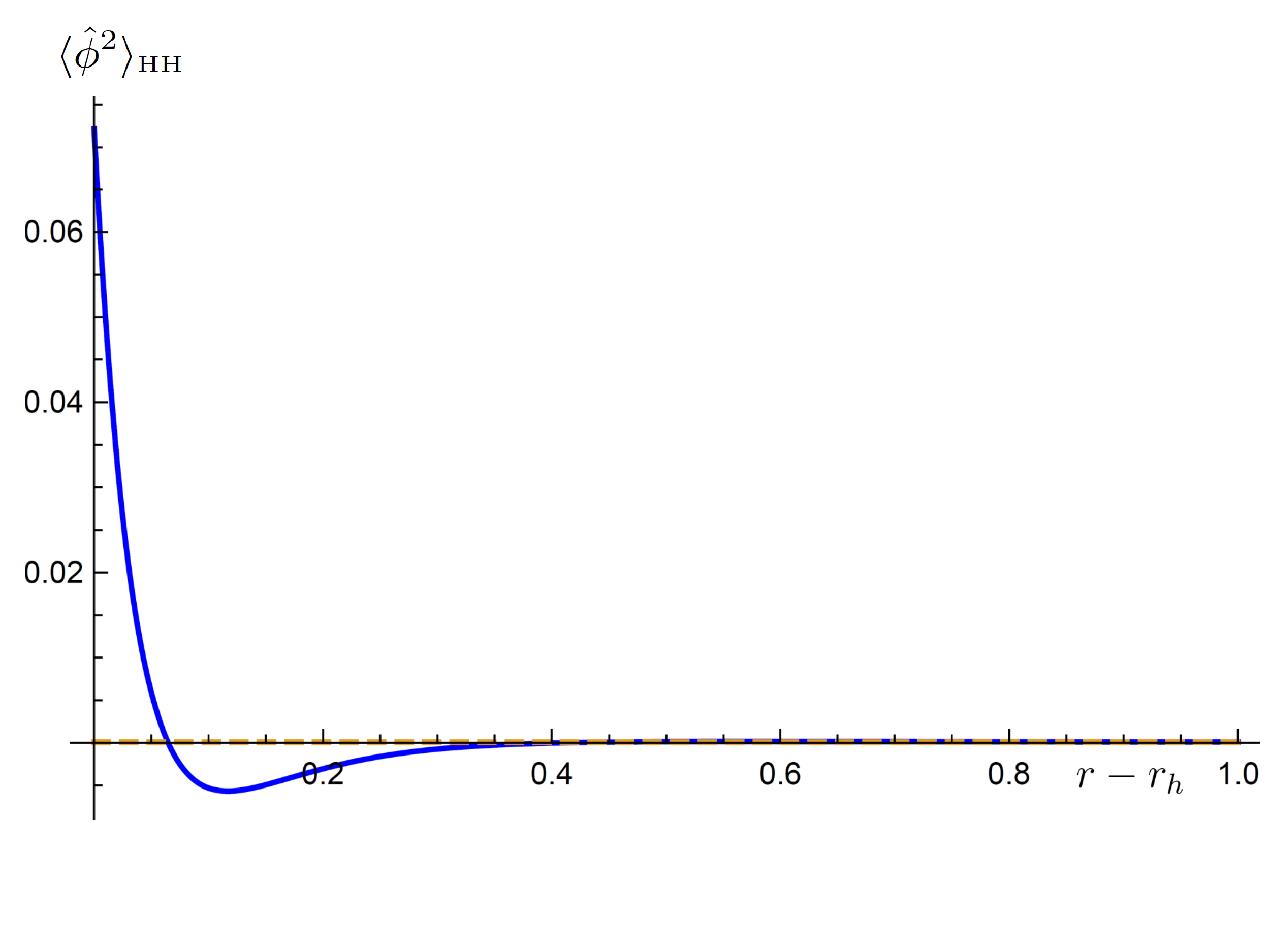}
		\caption{$d=8$}
		\label{fig:d8}
	\end{subfigure}%
	\begin{subfigure}{.5\textwidth}
		\includegraphics[width=.8\linewidth]{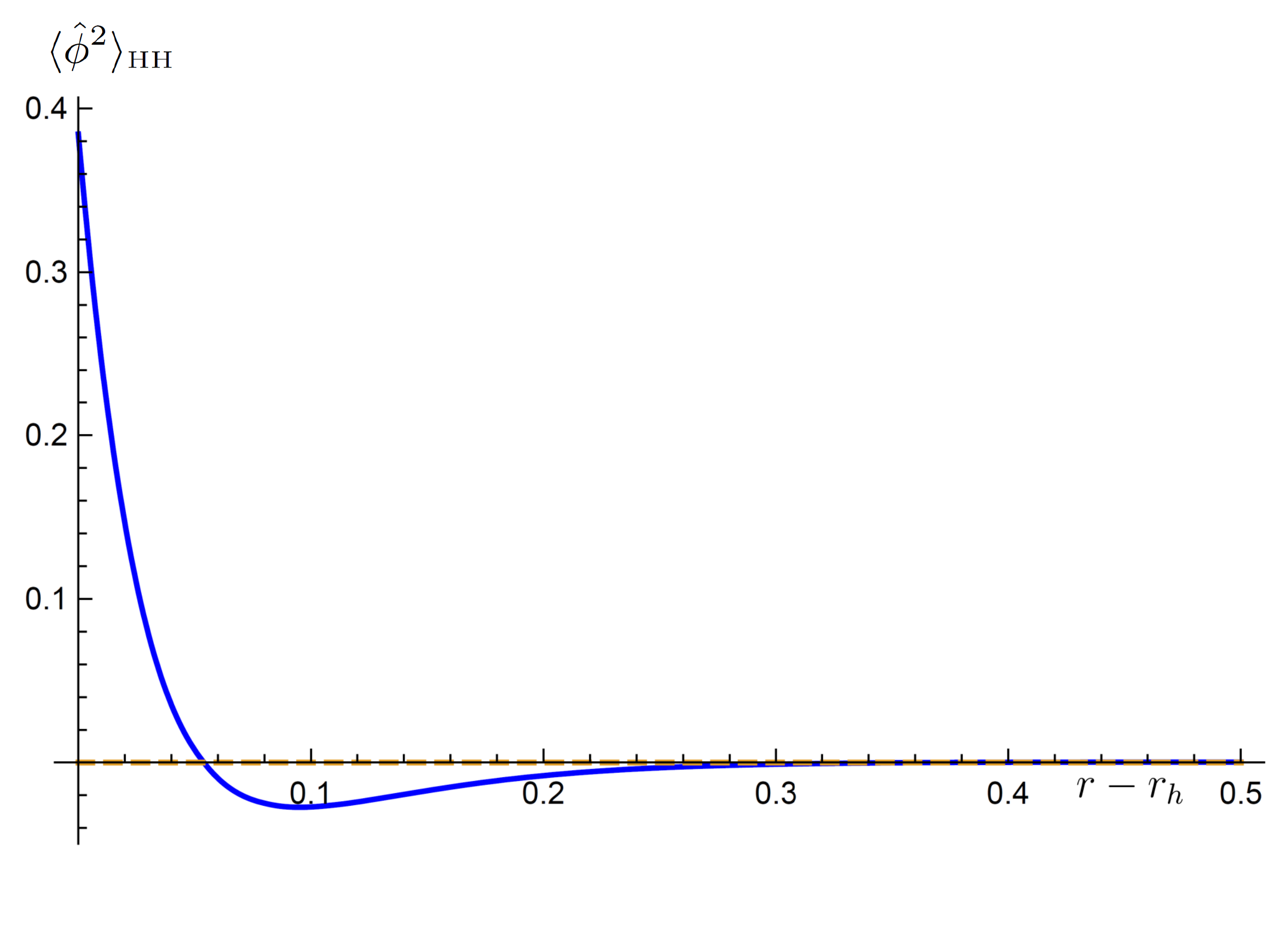}
		\caption{$d=9$}
		\label{fig:d9}
	\end{subfigure}
	\caption{\label{fig:dind}Individual plots of the vacuum polarization for a scalar field in SadS spacetime for $d=4$ to $d=9$ inclusive, with parameter values $\varpi_{d}=2,~m=0$ and $\xi=1/6$. In each of these plots the dashed line represents the value of the vacuum polarization for a massless scalar field in pure adS spacetime  with $\xi=1/6$. In each case, we see that the vacuum polarization approaches the adS value as $r\to\infty$. }
\end{figure*}

 In Fig.~\ref{fig:dind} we present individual plots for each spacetime dimension under consideration ($d=4...9$). We also include the constant pure adS values for comparison in this plot. We see that, in both the even and the odd $d$ case, the magnitude of $\langle \hat{\phi}^{2} \rangle_{\subHH}$ in the vicinity of the black hole horizon increases with increasing $d$. Moreover, the rate of change seems to be greater and the turning point closer to the horizon as the number of dimensions is increased. These graphs are qualitatively similar to those in the asymptotically flat Schwarzschild case \cite{taylorbreen:2016,taylorbreen:2017}, except that the graphs asymptote to the constant vacuum polarization for a  scalar field in adS rather than Minkowski spacetime.

\subsubsection{Asymptotic and Horizon Values}

 As mentioned above, the vacuum polarization for SadS ought to asymptote to the pure adS value for $r\to\infty$, and the rate of convergence to this value should be faster in higher numbers of dimensions as a result of the metric function Eq. (\ref{eq:fSadS}). In particular for the quantum field in the Hartle-Hawking state in SadS, given that the local measured temperature in SadS vanishes as $r \to \infty$ \cite{Hawking:1982}, the vacuum polarization should asymptote to the vacuum polarization for the natural vacuum state in adS.  We have explicitly verified this by checking our results against the explicit calculation of the vacuum polarization for a scalar field in the natural (zero temperature) vacuum state in adS in Ref. \cite{winstanleykent:2014}. Moreover, it is clear that the rate at which $\langle \hat{\phi}^2 \rangle_{\subHH}$ approaches this asymptotic adS value increases with $d$, as expected.

Finally as an additional check of our method in the $d=4$ case, we may use the expression for $\langle \hat{\phi}^2 \rangle_{\subHH}$ for a scalar field on the event horizon of any static spherically symmetric black hole spacetime derived in \cite{breenottewill:10}, and we find that for each set of parameters, our numerically calculated value just off the horizon matches up smoothly with the known horizon value. 

\subsection{ Approximations}
Although the method we presented in this paper is extremely efficient compared with other prescriptions for computing the vacuum polarization, it is nevertheless a difficult calculation because of the need to regularize the two-point function and because, in general, the modes must be calculated numerically. As a result, it is often useful to have approximations for the vacuum polarization and stress-energy tensor. An oft-used approximation for massive fields is the DeWitt-Schwinger approximation (see, for example, \cite{Matyjasek:2018,Thompson:2009}). Based on heat kernel methods for expanding the singular field, it is a purely local approximation independent of the quantum state. For $d=4$, the DeWitt Schwinger approximation is
\begin{align}
	\label{eq:DS4}
	\langle \hat{\phi}^2 \rangle_{\subDS}= \frac{1}{16\pi^2 m^2}\bigg[ \frac{1}{2}\left(\xi-\frac{1}{6}\right)^2R^2-\frac{1}{6}\left(\xi-\frac{1}{5}\right) \Box R \nonumber\\
	 +\frac{1}{180}\left(R_{abcd}	R^{abcd}	-R_{ab}	R^{ab}\right)\bigg]
\end{align}
while for $d=5$, we have
\begin{align}
	\label{eq:DS5}
	\langle \hat{\phi}^2 \rangle_{\subDS}= \frac{1}{32\pi^2 m}\bigg[ \frac{1}{2}\left(\xi-\frac{1}{6}\right)^2R^2-\frac{1}{6}\left(\xi-\frac{1}{5}\right) \Box R \nonumber\\
	+\frac{1}{180}\left(R_{abcd}	R^{abcd}	-R_{ab}	R^{ab}\right)\bigg]
\end{align}
where $R_{abcd}$ is the Riemann tensor and $R_{ab}$ is the Ricci tensor \cite{Matyjasek:2018,Thompson:2009}. The approximation (\ref{eq:DS4}) has been shown to be very accurate in the region of the black hole event horizon for $m\,M \geq 2$ in asymptotically flat black hole spacetimes \cite{Anderson:1990}. However, as is immediately obvious from inspection of (\ref{eq:DS4}) this approximation tends to $0$ as $m \to \infty$ in stark contrast to the large mass behaviour observed in the exact numerical calculation discussed in Section \ref{sec:m}. Hence, (\ref{eq:DS4}) completely fails as an approximation for the the vacuum polarization in SadS spacetime. A new approximation is required, which we will now derive. 

As the large mass behaviour appears to be a consequence of the fact that the spacetime is asymptotically adS, a natural starting point for a large mass approximation would be the closed form expressions for the vacuum polarization in a pure adS spacetime obtained in \cite{winstanleykent:2014}, which for $d=4$ and $d=5$ are given by
\begin{align}
	\label{eq:Phi2ads4}
	\langle \hat{\phi}^2 \rangle_{\textrm{adS}}=\frac{1}{8\pi^2 L^2}\bigg[\mu_\xi\bigg\{\psi\left(\sqrt{\mu_\xi+\frac{1}{4}} +\frac{1}{2}\right)+\gamma \nonumber\\
	-\log\left(2 L/\ell\right)\bigg\}-\frac{\mu_\xi}{2}-\frac{1}{6} \bigg]
	\end{align}
and
\begin{align}
	\langle \hat{\phi}^2 \rangle_{\textrm{adS}}=\frac{1}{24 \pi^2 L^3} \left(\frac{(4\mu_\xi-3)\sqrt{4\mu_\xi+1}}{8}\right),
	\end{align}
respectively. In the expression for $d=4$, $\psi(z)$ is the digamma function and the parameter $\ell$ is again an arbitrary lengthscale needed to make the $\log$ term dimensionless. Here $\mu_{\xi}$ is the effective dimensionless mass defined by (\ref{eq:muxi}).

Starting with the $d=4$ case, we wish to add to (\ref{eq:Phi2ads4}) terms that incorporate a dependence on the black hole mass. We note that the DeWitt-Schwinger approximation would contribute terms that depend on $L$, which we assume have already been accounted for by the adS term (\ref{eq:Phi2ads4}), and also a term like
\begin{align}
	\frac{M^{2}}{60\pi^{2}m^{2}r^{6}}.
\end{align}
Rather than simply adding this expression, we recall that the vacuum polarization depends not on the mass but on the effective mass $\mu_{\xi}$. Moreover, an approximation in terms of $\mu_{\xi}$ would be valid for massless fields. So we substitute $m^{2}\to (\mu_{\xi}+\alpha)/L^{2}$, for some dimensionless $\alpha$ which we determine by comparing with the numerical results. In turns out that for (almost) all values of the field mass and coupling constant that we checked, $\alpha=3$ yields an extremely accurate approximation over the entire exterior black hole spacetime. Interestingly, this choice of $\alpha$ corresponds to the replacement $m^{2}\to \mu_{\xi}/L^{2}-\Lambda$. Hence, we have for $d=4$,
 \begin{align}
 	\langle \hat{\phi}^2 \rangle_{\subHH}\approx\frac{1}{8\pi^2 L^2}\bigg[\mu_\xi\bigg\{\psi\left(\sqrt{\mu_\xi+\frac{1}{4}} +\frac{1}{2}\right)+\gamma \nonumber\\
 	-\log\left(2 L/\ell\right)\bigg\}-\frac{\mu_\xi}{2}-\frac{1}{6} \bigg]+\frac{M^2 L^{2}}{60 \pi ^2 (\mu_{\xi}+3)r^6}.
 \end{align}

Turning now to the $d=5$ case. Applying the same arguments as before, we add to the pure adS value the $M$-dependent term from the DeWitt-Schwinger approximation, which in this case is
\begin{align}
	\frac{4\,M^2}{45 \pi ^4 m r^8}.
\end{align}
Next we again make the replacement $m^{2}\to (\mu_{\xi}+\alpha)/L^{2}$ and compare with the exact numerical plots to determine $\alpha$. In this case we find that $\alpha=6$ gives excellent agreement which again corresponds to the replacement rule $m^{2}\to \mu_{\xi}/L^{2}-\Lambda$, since $L=\sqrt{-(d-1)(d-2)/(2 \Lambda)}$. The result is that, for $d=5$, we have
\begin{align}
\langle \hat{\phi}^2 \rangle_{\subHH}\approx\frac{1}{24 \pi^2 L^3} \left(\frac{(4\mu_\xi-3)\sqrt{4\mu_\xi+1}}{8}\right) \nonumber\\
+\frac{4\,M^{2}L}{45 \pi ^4 \sqrt{\mu_{\xi}+6} r^8}.
\end{align}
We conjecture then that in all numbers of dimensions, an excellent approximation can be obtained by adding to the adS value for vacuum polarization the $M$-dependent term from the leading-order DeWitt-Schwinger approximation with the replacement rule $m^{2}\to \mu_{\xi}/L^{2}-\Lambda$.

\begin{figure*}
	\begin{subfigure}{.5\textwidth}
		\includegraphics[width=.8\linewidth]{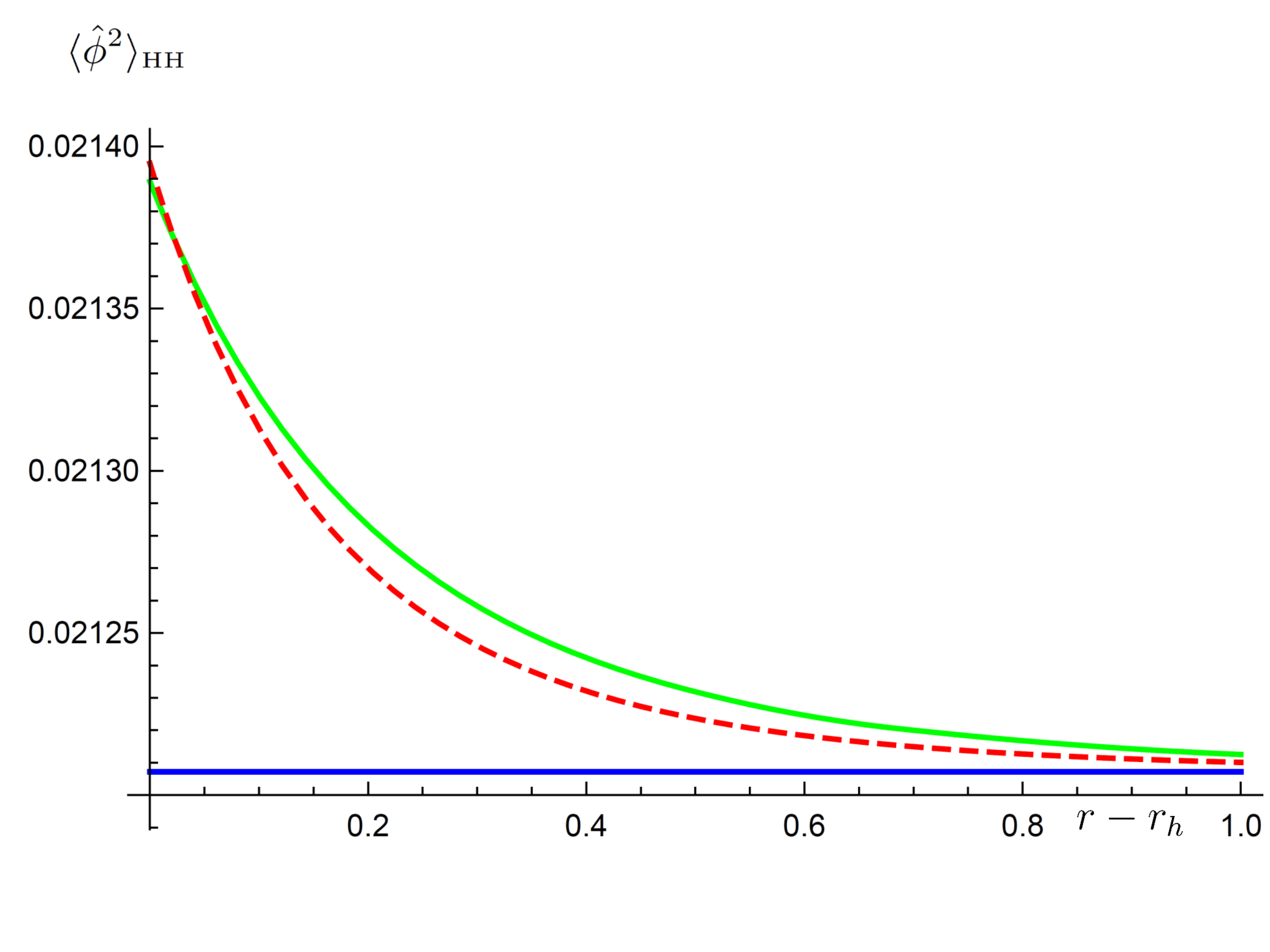}
		\caption{$d=4,m=2,\xi=0$, max error $\approx 0.06 \%$}
		\label{fig:d4approxm2}
	\end{subfigure}%
	\begin{subfigure}{.5\textwidth}
		\includegraphics[width=.8\linewidth]{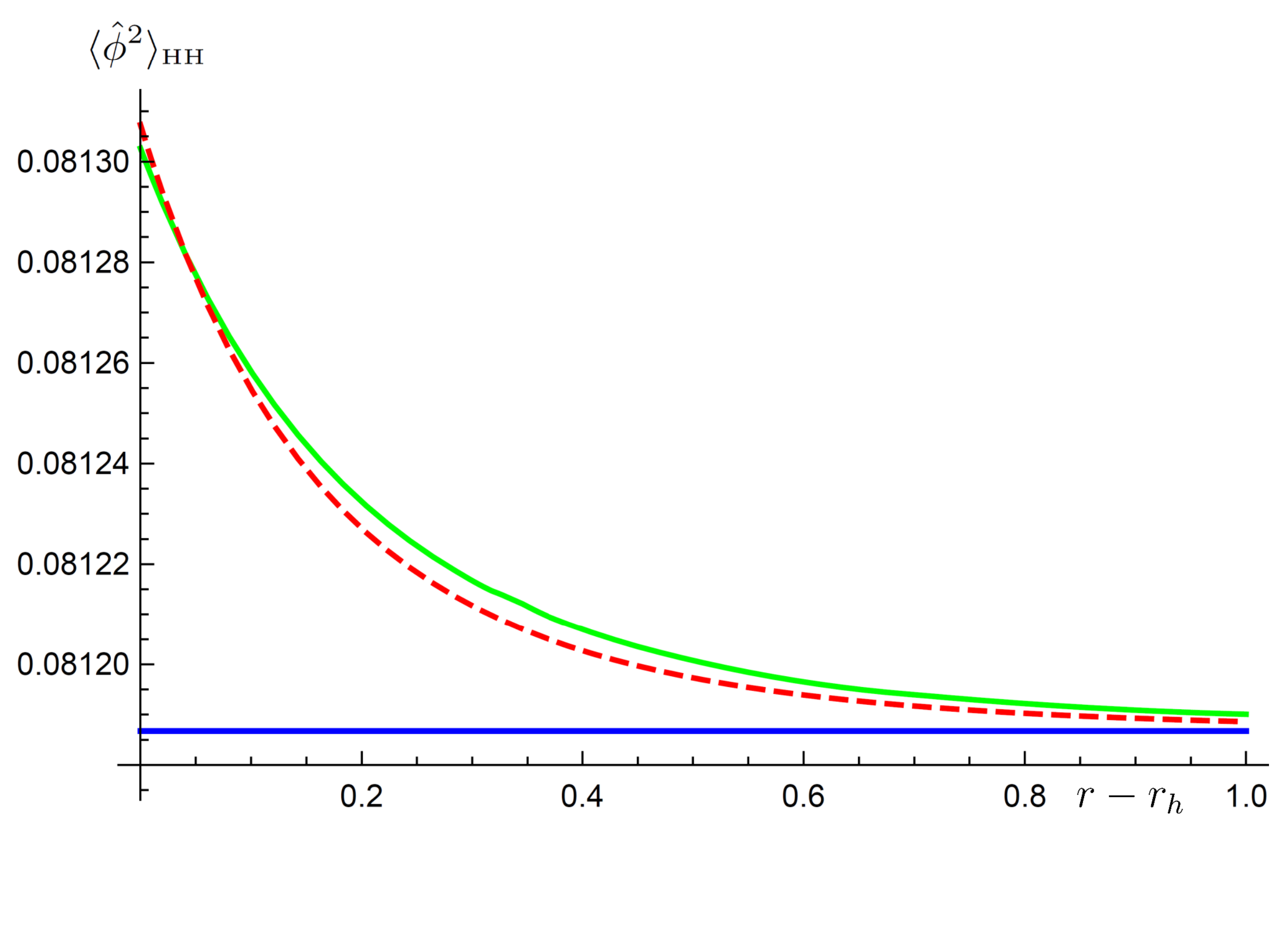}
		\caption{$d=4,m=3,\xi=0$, max error $\approx 0.007\%$}
		\label{fig:d4approxm3}
	\end{subfigure}
	\begin{subfigure}{.5\textwidth}
		\includegraphics[width=.8\linewidth]{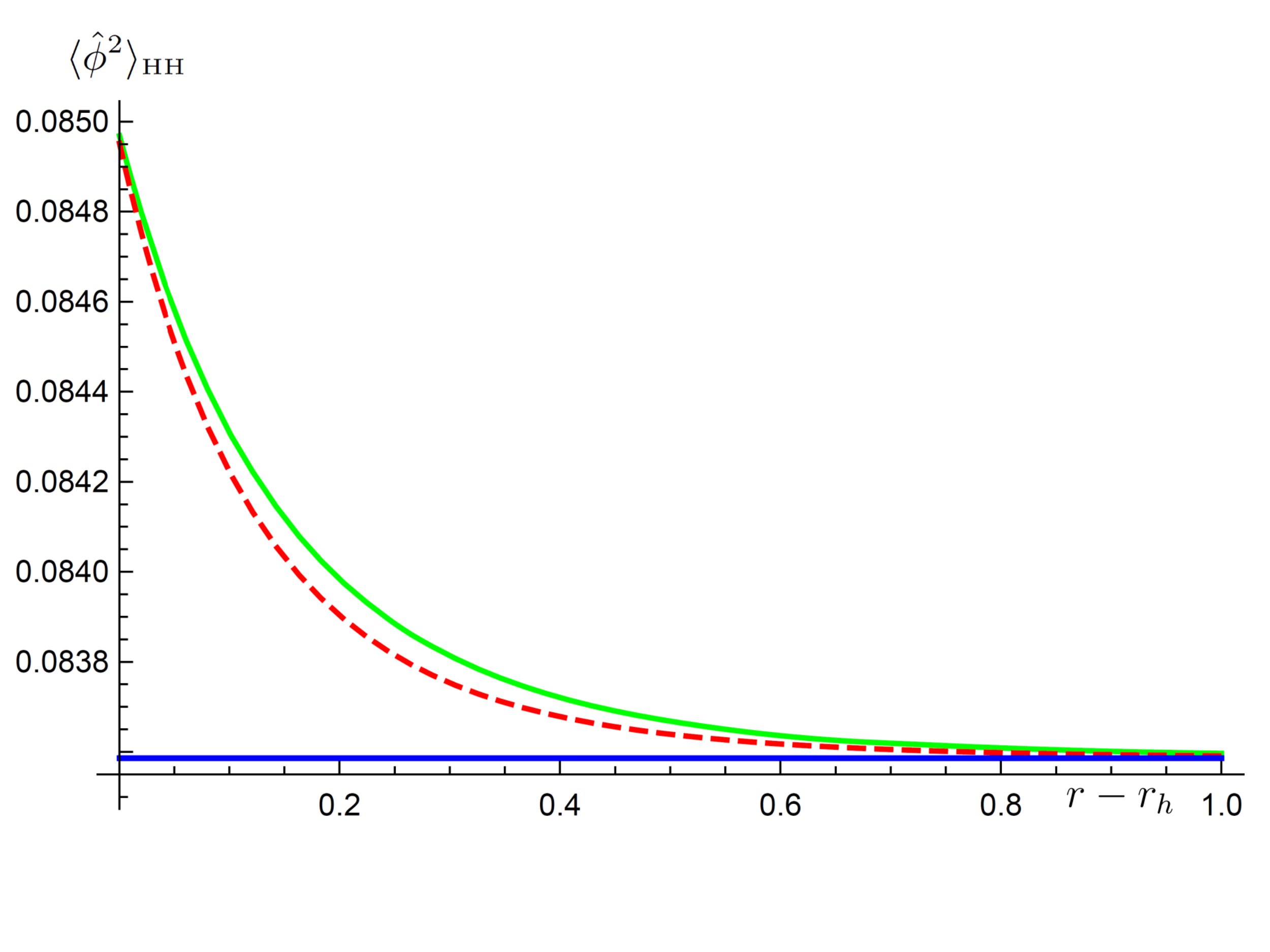}
		\caption{$d=5, m=2,\xi=0$, max error $\approx 0.1 \%$}
		\label{fig:d5approxm2}
	\end{subfigure}%
	\begin{subfigure}{.5\textwidth}
		\includegraphics[width=.8\linewidth]{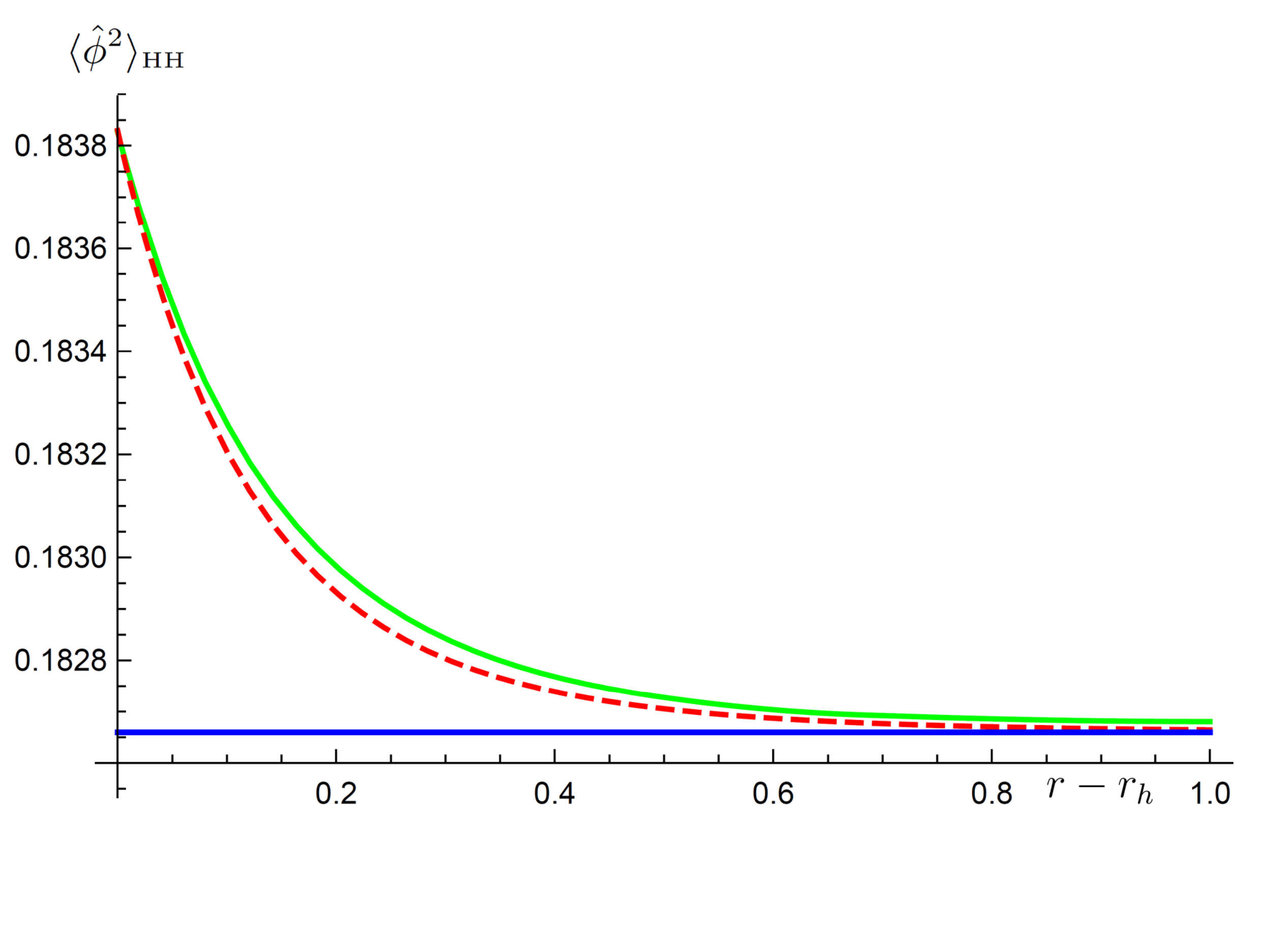}
		\caption{$d=5, m=3,\xi=0$, max error $\approx 0.03 \%$}
		\label{fig:d5approxm3}
	\end{subfigure}
	\caption{\label{fig:approxm} Plots of the vacuum polarization alongside our approximation scheme, denoted by the dashed line, for a minimally coupled scalar field in SadS spacetime for $d=4$ and $d=5$ with field mass $m=2$ and $m=3$. We have fixed $L=\ell=1$ and $\varpi_{d}=2$ (corresponding to $M=1$ and $M=3\pi/4$ for $d=4$ and $d=5$, respectively). In each plot the constant function represents the pure adS vacuum polarization.}	
\end{figure*}
\begin{figure*}
	\begin{subfigure}{.5\textwidth}
		\includegraphics[width=.8\linewidth]{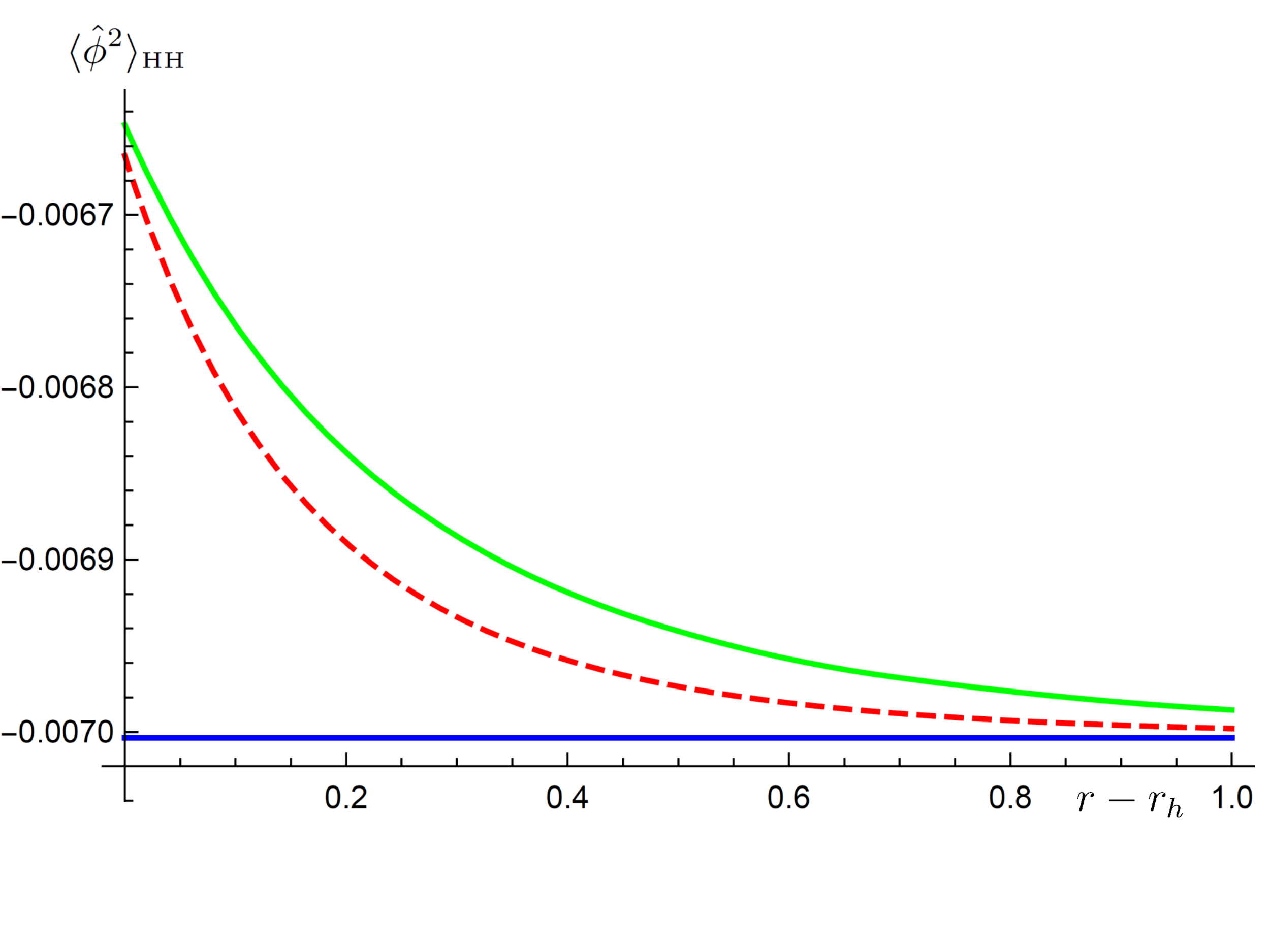}
		\caption{$d=4,m=0,\xi=0$,  max error $\approx 0.8 \%$}
	\label{fig:d4approxxim16}
	\end{subfigure}%
	\begin{subfigure}{.5\textwidth}
		\includegraphics[width=.8\linewidth]{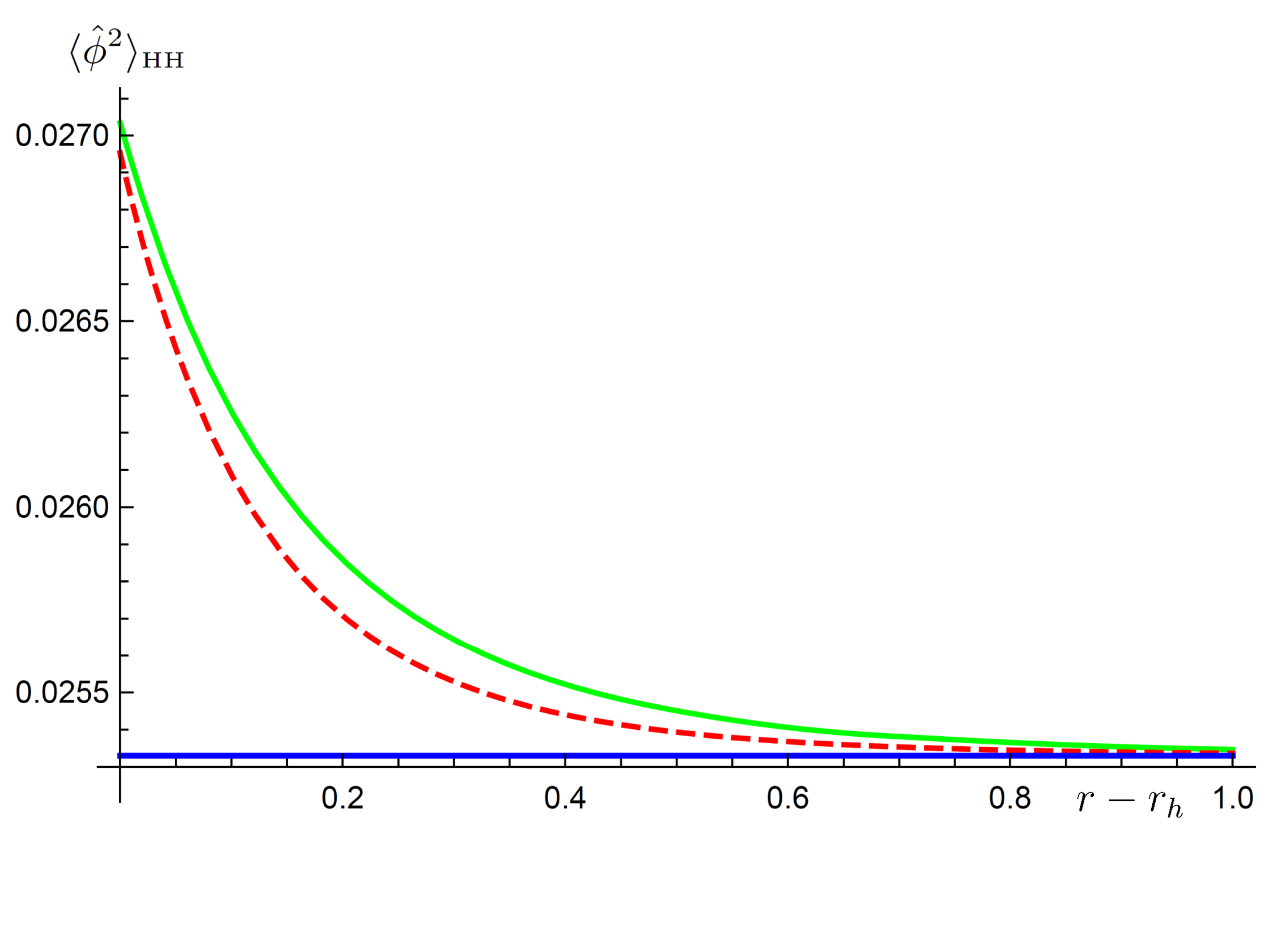}
		\caption{$d=5,m=0,\xi=0$,  max error $\approx 0.7 \%$}
		\label{fig:d5approxxim16}
	\end{subfigure}
	\caption{\label{fig:approxxi} Plots of the vacuum polarization alongside our approximation, denoted by the dashed line, for a massless scalar field in SadS spacetime for $d=4$ and $d=5$. The coupling constant is $\xi=-1/6$. We have set $L=\ell=1$ and $\varpi_{d}=2$. In each plot the constant function represents the adS vacuum polarization.}	
\end{figure*}

In Figs. ~(\ref{fig:approxm}) and (\ref{fig:approxxi}), the comparison of our new approximation to our exact numerical results is presented. Even though the $M$-dependent terms in our approximation were derived from the large $m$ limit, it can be seen to be an excellent approximation to the exact numerical results for almost all values of the mass and coupling constant over the entire exterior spacetime in both the $d=4$ and the $d=5$ cases, with maximum errors of less than $1\%$, even in the massless (non-conformally coupled) case. It is clear from the plots that the approximation is most accurate near the horizon and for large $r$ values, however the error in the intermediary region is still extremely small. The accuracy of the approximation increases with increasing $\mu_\xi$ and decreases significantly as $\mu_\xi \to 0$ (massless, conformally coupled case), with a max error in this case of $\approx 25\%$.  We note that one of the reasons that this approximation is so successful is that it is not a local approximation, but rather a global one in the sense that it depends on the quantum state via the exact adS values for vacuum polarization that we used as our starting point for the approximation. Another advantageous feature of the approximation is that it has the correct renormalization ambiguity in even numbers of dimensions, encoded in it the $\log(\ell)$ term. The coefficient in front of this term is precisely $v_{0}$, the coincidence limit of the Hadamard biscalar $V(x,x')$. This corresponds to the freedom to add to our singular Hadamard parametrix homogeneous solutions of the wave equation.

 As a final note in this section, we conjecture that the approximation scheme we have outlined here can also be applied to the renormalized expectation value of the stress-energy tensor in SadS spacetimes, given that exact expressions for the stress energy tensor in adS have also been calculated in \cite{winstanleykent:2014} for $3\leq d \leq 11$. Of course, in order to test the validity of any such approximation, a full numerical calculation for the renormalized expectation value of the stress-energy tensor in an asymptotically adS higher dimensional black hole spacetime would be required, we hope to report on this in the future.

\section{Conclusions}
\label{sec:con}
In this paper, we have conducted a detailed study of the dependence of quantum vacuum polarization effects due to a scalar field on the parameters of the field mass and coupling and on the parameters of the background black hole spacetime. Most calculations of this type in the literature focus on a single fixed set of parameters for the field and the background spacetime, due to the complexity and inefficiency of the usual methods used to renormalize the vacuum polarization. We adopt a very recent method, which we call the extended coordinate method, for computing the vacuum polarization in static spherically-symmetric spacetimes. This method is extremely efficient and provides a rapidly convergent mode-sum expression for the vacuum polarization for arbitrary field parameters, arbitrary metric function $f(r)$ and arbitrary numbers of dimensions. The robustness and efficiency of this method permits a detailed analysis of the dependence of quantum effects on the various parameters in the theory. 

Of particular interest is the dependence on the field mass and scalar coupling constant, which show features very different from the vacuum polarization in asymptotically flat black holes. Considering the dependence on the field mass first, we found that the vacuum polarization depends only weakly on the field mass for small values. Hence, it is probably reasonable to consider a massless field approximation even when the field mass is small but non-zero. When the field mass is large however we see that the vacuum polarization can become arbitrarily positive for $d=4,5$ or arbitrarily negative for $d=6,7$, a fact which has serious implications for the validity of the usual DeWitt-Schwinger approximation. This behaviour  is an artefact of the asymptotic structure of the spacetime under consideration  , a fact which we exploited to develop a new and extremely accurate approximation to the vacuum polarization, valid for all values of the field mass. We further conjectured an approximation method for the renormalized stress-energy tensor. The dependence of the vacuum polarization on the coupling constant and the field mass is through an effective dimensionless mass (\ref{eq:muxi}). This effective mass is degenerate for distinct pairs of mass and coupling parameters. For example, the vacuum polarization for a minimally coupled massive field can be identical to that of a massless non-minimally coupled field. In either case, the vacuum polarization can be arbitrarily large for sufficiently negative values of the coupling, or equivalently, for sufficiently large field mass, even without violating any of the approximations underlying the semi-classical picture. We showed that this behaviour is an artefact of the Dirichlet boundary conditions we have adopted. For Neumann and Robin boundary conditions, well-posedness of the wave equation constrains the effective mass to be within $-1/4<\mu_{\xi}<3/4$ which provides both upper and lower bounds for the coupling and field mass. This presumably cures any potentially large back-reaction effects arising from large field mass or large negative couplings. However, a detailed comparison of the renormalized stress-energy tensor and back reaction on the SadS background for different boundary conditions would offer some insights on this matter.

\bibliographystyle{apsrev4-1}
%


\end{document}